\documentclass[
 aps, pra,
 amsmath,amssymb,
 11pt,
 final,
tightenlines,
 twoside,
 twocolumn,
 nofloats,
nofootinbib,
 superscriptaddress,
showkeys,
showkeywords,
 ]
{revtex4-2}

\usepackage[T2A]{fontenc}
\usepackage[utf8x]{inputenc}
\usepackage[english]{babel}
\usepackage{graphicx}
\usepackage{dcolumn}
\usepackage{bm}
\usepackage{longtable}
\usepackage{MnSymbol}

\usepackage{multirow}

\usepackage[dvipsnames]{xcolor}
\definecolor{myred}{rgb}{0.90, 0.1, 0.1}
\definecolor{myblue}{rgb}{0.1, 0.1, 0.90}
\definecolor{mygreen}{rgb}{0.1, 0.90, 0.1}
\definecolor{mygray}{gray}{0.6}

\input{maik.rty}

\setcitestyle{authoryear,round,aysep={,},citesep={;}}
\def\squareforqed{\hbox{\rlap{$\sqcap$}$\sqcup$}}

\def\sq{\ifmmode\squareforqed\else{\unskip\nobreak\hfil
\penalty50\hskip1em\null\nobreak\hfil\squareforqed
\parfillskip=0pt\finalhyphendemerits=0\endgraf}\fi}

\def\utw{\smash{\rlap{\lower5pt\hbox{$\sim$}}}}

\def\udtw{\smash{\rlap{\lower6pt\hbox{$\approx$}}}}

\def\diameter{{\ifmmode\mathchoice
{\ooalign{\hfil\hbox{$\displaystyle/$}\hfil\crcr
{\hbox{$\displaystyle\mathchar"20D$}}}}
{\ooalign{\hfil\hbox{$\textstyle/$}\hfil\crcr
{\hbox{$\textstyle\mathchar"20D$}}}}
{\ooalign{\hfil\hbox{$\scriptstyle/$}\hfil\crcr
{\hbox{$\scriptstyle\mathchar"20D$}}}}
{\ooalign{\hfil\hbox{$\scriptscriptstyle/$}\hfil\crcr
{\hbox{$\scriptscriptstyle\mathchar"20D$}}}}
\else{\ooalign{\hfil/\hfil\crcr\mathhexbox20D}}%
\fi}}

\begin{document}

\selectlanguage{english}



\title{Search and Study of the Brightest Stars in the Galaxy IC\,342}

\author{\firstname{O.~N.}~\surname{Sholukhova}}
\email{olga@sao.ru}
\affiliation{Special Astrophysical Observatory,  Russian Academy of Sciences, Nizhnii Arkhyz, 369167 Russia}
\author{\firstname {N.~A.}~\surname{Tikhonov}}
\affiliation{Special Astrophysical Observatory,  Russian Academy of Sciences, Nizhnii Arkhyz, 369167 Russia}

\author{\firstname{Y.~N.}~\surname{Solovyeva}}
\affiliation{Special Astrophysical Observatory,  Russian Academy of Sciences, Nizhnii Arkhyz, 369167 Russia}

\author{\firstname{A.~N.}~\surname{Sarkisian}}
\affiliation{Special Astrophysical Observatory,  Russian Academy of Sciences, Nizhnii Arkhyz, 369167 Russia}
 
\author{\firstname{A.~S.}~\surname{Vinokurov}}
\affiliation{Special Astrophysical Observatory,  Russian Academy of Sciences, Nizhnii Arkhyz, 369167 Russia}

\author{\firstname{A.~T.}~\surname{Valcheva}}
\affiliation{Sofia University  St. Kliment Ohridski,  Sofia,  1164 Bulgaria}

\author{\firstname{P.~L.}~\surname{Nedialkov}}
\affiliation{Sofia University  St. Kliment Ohridski,  Sofia,  1164 Bulgaria}

\author{\firstname{D.~V.}~\surname{Bizyaev}}
\affiliation{Apache Point Observatory and New Mexico State University, Sunspot,  88349-0059 USA}
\affiliation{Sternberg Astronomical Institute, Moscow State University, Moscow, 119234 Russia}

\author{\firstname{B.~F.}~\surname{Williams}}
\affiliation{ University of Washington, Seattle, 98195 USA}

\author{\firstname{V.~D.}~\surname{Ivanov}}
\affiliation{European Southern Observatory,  Garching, D-85748 Germany}

\begin{abstract}
We have selected candidate massive stars in the galaxy IC\,342 based on archival images from the Hubble Space Telescope and images from the 2~m telescope at the Natioal Astronomical Observatory Rozhen, Bulgaria. Spectral observations of 24 out of 27 selected stars are carried out with the 6~m BTA telescope at the SAO RAS and with the 3.5\,m Apache Point Observatory telescope (USA) as part of the program for searching bright massive stars in galaxies outside the Local Group. Our analysis reveals that 12~objects have spectra lacking prominent features, except for the emission lines of the surrounding nebulae and are identified as single supergiants of classes O9 to F5 or spatially unresolved young compact clusters. One source with an absorption spectrum probably belongs to our Galaxy. The spectra of seven other objects show features typical of Wolf-Rayet stars or compact clusters containing Wolf--Rayet stars. Another source is a compact supernova remnant. Two other objects are tentatively classified as cold LBV candidates, and one object is classified as a B[e]-supergiant candidate.
\end{abstract}

\maketitle

\section{INTRODUCTION}

One of the main tasks of modern astrophysics is to study the evolution of the most massive stars. Throughout their lives, these stars have very high luminosities, populating the upper part of the Hertzsprung-Russell diagram. The lifetime of these stars is limited to 3--5\,Myr, with final evolutionary stages potentially resulting in a supernova explosion and the formation of a black hole or a neutron star. Despite active research in this area, many questions remain unanswered.

One such question concerns the upper limit on the mass of newly born stars. It is assumed that objects belonging to the so-called very massive stars (VMS) with masses reaching several hundred solar masses can be formed in the cores of massive star clusters through the dynamic mergers of several lower mass stars (Portegies
Zwart et al., 1999). 
At the same time, a significant fraction of massive stars in binary systems can be kicked out of the cluster as a result of 3--4-fold collisions (interactions) of stars (Oh and Kroupa, 2016), therefore they can often be observed near parent associations (Kostenkov et al., 2017; Figer
et al., 2020). There is also evidence for the formation of massive stars outside stellar associations. For example, Oskinova et al. (2013) have not found any connections between star in our Galaxy WR\,102ka with a recent mass of approximately 100\,$M_\odot$ (the initial mass was about 150\,$M_\odot$) with the surrounding stellar clusters  (Oskinova et al., 2013).

After the onset of thermonuclear reactions, the evolution of massive stars is significantly influenced by the mass loss through stellar winds; the rate of this loss can increase dramatically after the stars leave the main sequence. The presence of powerful winds simplifies the search for such stars, because emission lines begin to dominate their spectra. 
Our search criteria for massive stars---identified as the brightest stars with H$\alpha$ emission, located inside or in the vicinity of young clusters or star-forming regions---yield a variety of different types of stars in the upper part of the Hertzsprung-Russell diagram. These types include luminous blue variables (LBVs), B[e] supergiants, Fe\,II stars  (Humphreys et al., 2017), yellow hypergiants (Klochkova, 2019), supernova impostors  (Pastorello and Fraser, 2019), as well as some bright WR stars or compact WR star clusters. The latter can be as small as a few parsecs in size, (Kurtev et al., 2007) which, at sufficiently large distances, makes them indistinguishable from single stars even in Hubble Space Telescope (HST) images.

Some sub-classes of luminous stars are extremely rare. 
Richardson and Mehner (2018) presented a list of LBV stars that includes 41 objects in our and in nearby galaxies. The review of Kraus (2019) yields nine B[e] supergiants in our Galaxy and 24 in others. Discovering such objects in the Local Group galaxies will increase the statistics and allow to study their features as a function of the metallicity.

The most accurate stellar parameters can be obtained for the closest and brightest of stars located in our Galaxy. However, young stars in the Milky Way are born in the thin disk, where most of the gas and dust are concentrated. Due to this mix, some of these stars are subjected to such a strong extinction that they are only visible in the infrared. For example, the apparent magnitude of the star \mbox{LBV\,1806-20}  \mbox{($M=130$--$200\,M_\odot$}, Eikenberry et al., 2004)  
 is estimated to be $V=35^{\rm m}$, inaccessible in the optical range for any existing telescope. Therefore, to alleviate the stellar extinction and broaden the search, we should look beyond the Galaxy and search for high-luminosity stars in others, yet relatively nearby galaxies (Sholukhova et al.,
2018; Solovyeva et al., 2019, 2021, 2023).  HST images enable the search for such stars in galaxies at distances up to 20~Mpc.

\begin{figure*}[]
\includegraphics[width=12cm]{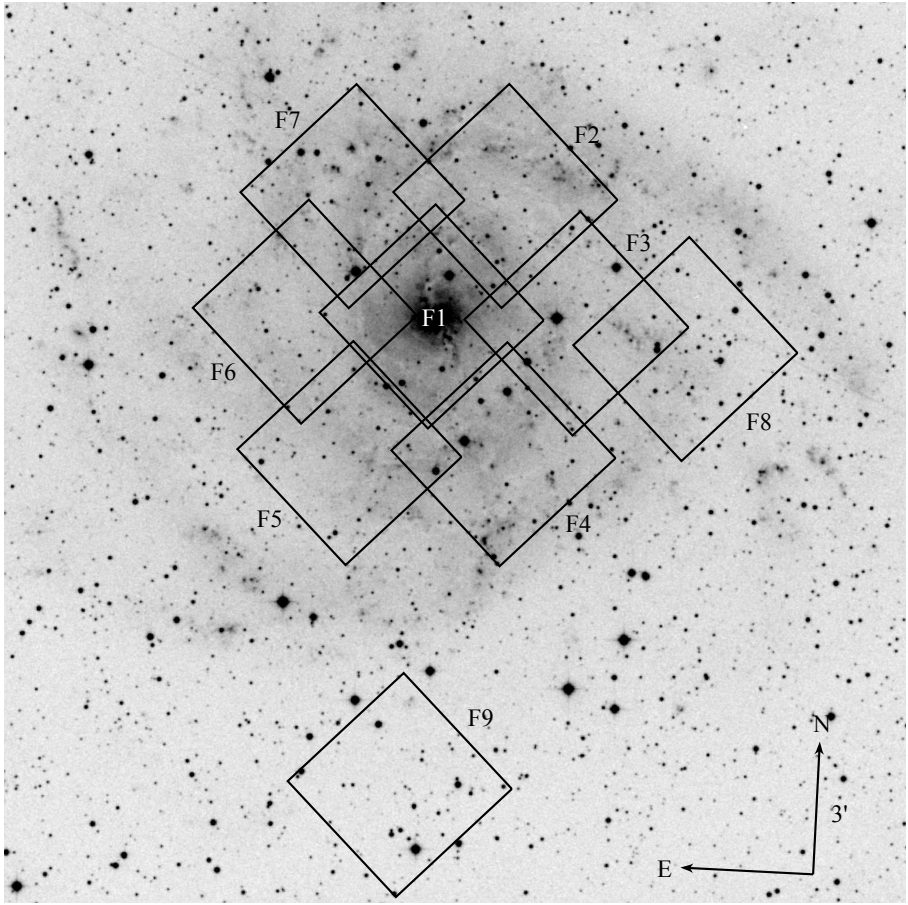} 
\caption{DSS image of IC\,342 galaxy. Nine HST fields are indicated.}
\label{fig:1}
\end{figure*}

Massive stars in dwarf galaxies are quite sparse  (Tikhonov et al., 2021b),  despite the high probability of forming such stars in low metallicity environments typical of low-mass galaxies (Hosokawa and Omukai, 2009). Since massive stars most often form in young star clusters, the search for them should focus on dwarf galaxies with active star formation. The star formation is continuously on-going in giant spirals, so the probability of finding a massive star in them is significantly higher than in dwarf galaxies. Spiral galaxies observed face-on are particularly interesting, because the entire galaxy is accessible to observations, and the stars are visible with minimal extinction by the gas and dust clouds in the disk. IC\,342, located at a distance of 3.9~Mpc, matches these requirements (Tikhonov and Galazutdinova, 2010, 2018). Although the extinction toward IC\,342 is quite significant  (Tikhonov and Galazutdinova, 2018, $A_V=1\,.\!\!^{\rm m}6$), the galaxy's proximity allows us to study in detail the structure of compact star clusters and to identify luminous stars in them.

The paper is organized as follows: Section~2 describes the methodology for selecting candidates and the photometric features of  HST images, and provides identification maps for the sample objects. Section~3 describes the spectroscopic data used in this work, including the observing log, and the processing stages. Section~4 provides examples of the spectra of the objects of interests and an analysis of all the studied stars. Section~5  briefly summarizes the main results of the work.

\section{STELLAR PHOTOMETRY AND SEARCH FOR CANDIDATE MASSIVE STARS}
\label{sec:phot}

\begin{figure}
\includegraphics[width=7.6cm]{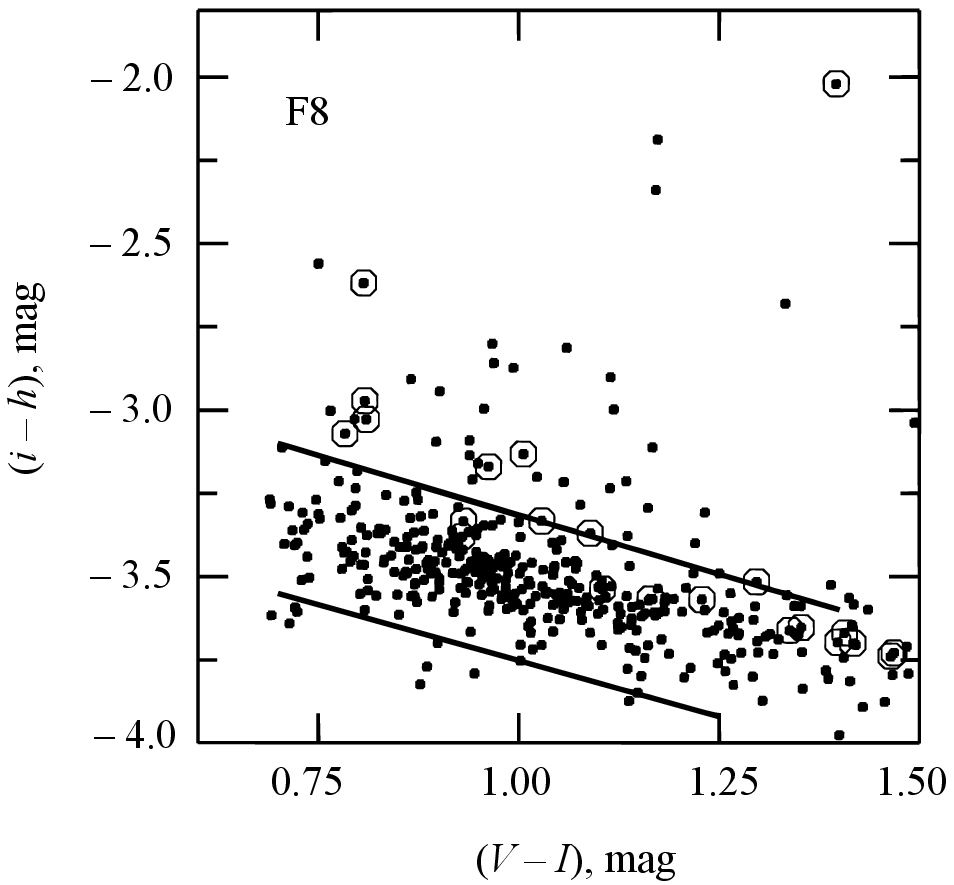}
\caption{Two-color diagram $(V-I)$--$(I-{\rm H}\alpha)$ of stars in the field F8. The color index ($I-{\rm H}\alpha)$ is given in relative units and is marked in the diagram as $(i-h)$. Stars with no H$\alpha$ emission are located between two parallel lines. Emission stars are located higher. Circles indicate bright stars ($V<21\,.\!\!^{\rm m}5$). The star F3.7 with the brightest H$\alpha$ line is located in the upper right corner of the plot.}
\label{fig:2}
\end{figure}

\begin{figure}
\includegraphics[width=7.7cm,clip]{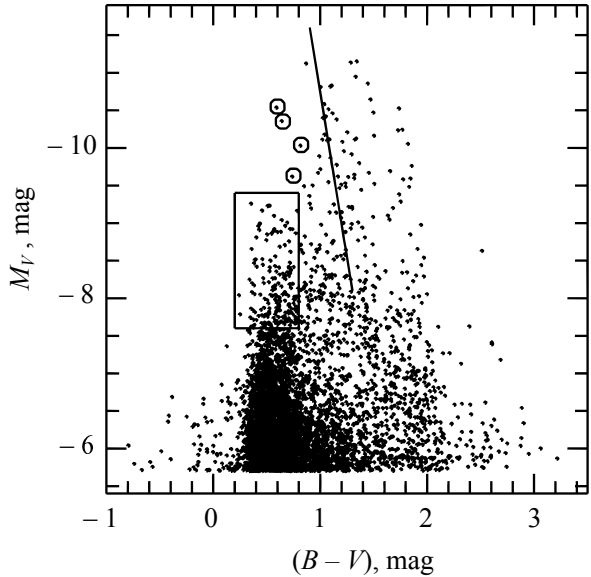}
\caption{Color--Magnitude diagram of bright stars in the fields F1--F8. The straight line shows the locus of the foreground Milky Way stars. The circles indicate bright blue semi-resolved objects in star clusters of IC\,342. The rectangle marks the locus of bright blue stars.}
\label{fig:3}
\end{figure}

\begin{figure*}
\includegraphics[width=0.96\textwidth]{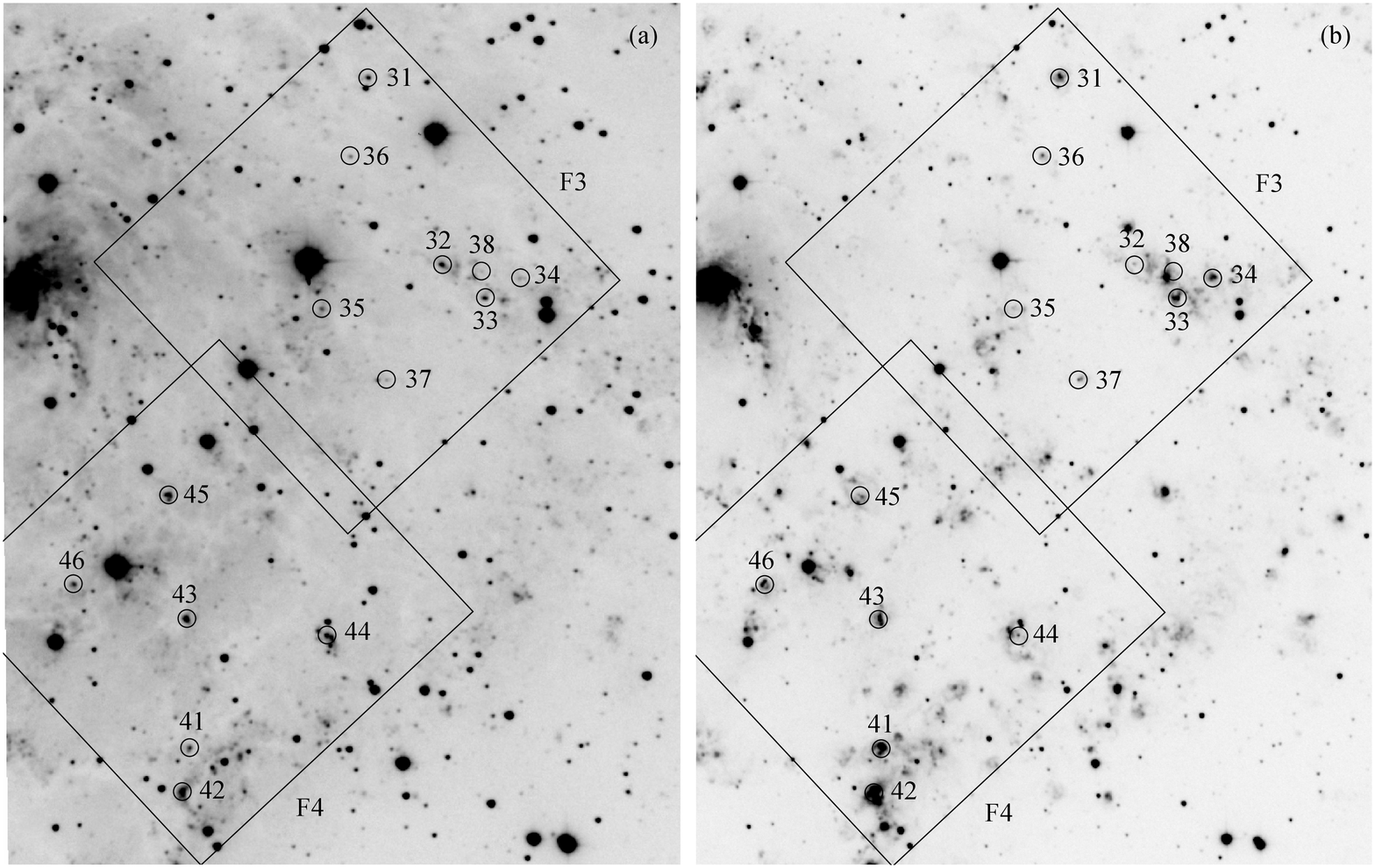}
\caption{Parts of the IC\,342 images, taken with the 2~m telescope at NAO Rozhen, Bulgaria, in $B$-band (a) and H$\alpha$ filter (b). The boundaries of F3 and F4 fields are shown, and the circles mark the locations of clusters containing selected bright blue stars.}
\label{fig:4}
\end{figure*}

To search for the brightest star candidates in the galaxy IC\,342, we used archival  HST images obtained under the proposals ID\,10768 and ID\,16002, as well as images from the 2~m telescope at the National astronomical observatory Rozhen (Bulgaria). The observations with the 2~m telescope were carried out using the focal reducer FoReRo2\footnote{\url{https://www.astro.bas.bg/forero/info/2KFR_info.htm}} in two nights: on October 10, 2019, images were taken in $B$ and H$\alpha$ filters, and on November 22, 2020 in $B$ and $R$ as part of the program for searching for luminous blue variables in galaxies outside the Local Group. Four IC\,342 fields were observed, each of size $10^\prime\times10^\prime$ with a small overlap. The final $B$, $R$ and H$\alpha$ images were obtained after the initial reduction and co-adding of nine (H$\alpha$) or five ($B$ and $R$) separate 300~s exposures. The seeing during the first night was about $1\,.\!\!^{\prime\prime}8$, and during the second---approximately $3^{\prime\prime}$.

Figure~1 shows DSS (Digital Sky Survey) image of IC\,342 with nine HST fields marked. For field F8, there were HST images in filters $F555W$ ($V$), $F814W$ ($I$) and $F658N$ (H$\alpha$), and for fields F1 to F7 and F9---in filters $F435W$ ($B$) and $F606W$ ($V$). The images of the 2-m telescope in the $B$, $R$ and H$\alpha$ filters completely covered the entire IC\,342 image shown in Fig.~1.

The photometry of stars was performed with the following software packages: \texttt{DAOPHOT~II} (Stetson, 1987, 1994) and \texttt{DOLPHOT~2.0} (Dolphin, 2016). The photometry in both cases used a standard approach. For \texttt{DAOPHOT~II}, we outlined the procedures earlier in Tikhonov at al. (2019), while \texttt{DOLPHOT~2.0} was employed following the guidelines provided by Dolphin (2016). The photometry involved preliminary masking of bad pixels, cosmic removal, and subsequent PSF photometry of the identified stars in two filters. To remove non-point objects, such as unresolved star clusters and distant or compact galaxies, all stars were selected based on their ``CHI'' and ``SHARP'' parameters, which characterize the shape of the photometric profile of each detected star (Stetson, 1987). The profiles of non-stellar objects differ from those of selected isolated stars, which enabled us to perform this selection using the star catalogs generated by both \texttt{DAOPHOT~II} and \texttt{DOLPHOT~2.0}. \texttt{DAOPHOT~II} was employed for the images of the field F8, while \texttt{DOLPHOT~2.0} was used for the images of F1--F7 and F9 fields.

\setlength{\tabcolsep}{3.5pt}
\renewcommand{\baselinestretch}{0.9}
\begin{table*}[!hbtp]
\caption{Massive star candidates in IC\,342. The typical error values of the quantities $V$, $(B-V)$, $A_V$, $(B-V)_0$, $M_V$ are $0\,.\!\!^{\rm m}02$, $0\,. \!\!^{\rm m}03$, $0\,.\!\!^{\rm m}09$, $0\,.\!\!^{\rm m}04$ and $0\, .\!\!^{\rm m}11$, respectively \medskip} 
\label{tab:star} 
\begin{tabular}{l|c|c|c|c|c|c|c|r@{$\,\pm\,$}l}
\hline
\multicolumn{1}{c|}{\multirow{2}{*}{ID}} & RA & Dec & $V$, & $(B-V)$, & $A_V$, &  $(B-V)_0$, & $M_V$, & \multicolumn{2}{c}{$M_{\rm bol}$,}\\
 & (J2000) & (J2000) & mag & mag & mag & mag & mag & \multicolumn{2}{c}{mag}\\
\hline 
F1.1 & 03 46 51.06 & +68 06 34.5 & 21.08 & 0.55 & 1.92 & $-0.07$ & \,$-8.81$ &  $ -9.40$ & $0.32$ \\
F2.1 & 03 46 33.89 & +68 10 05.9 & 21.80 & 0.48 & 2.36 & $-0.28$ & \,$-8.53$ &  $-11.76$ & $1.04$ \\
F2.2 & 03 46 39.00 & +68 09 32.8 & 19.95 & 0.68 & 2.14 & $-0.01$ & $-10.16$ &  $-10.48$ & $0.17$ \\
F2.3 & 03 46 22.34 & +68 08 58.9 & 20.69 & 0.45 & 1.98 & $-0.19$ & \,$-9.26$ &  $-11.15$ & $0.55$ \\
F2.4 & 03 46 49.28 & +68 07 37.4 & 20.87 & 0.56 & 1.98 & $-0.08$ & \,$-9.08$ &  $ -9.74$ & $0.30$ \\
F3.1 & 03 46 13.20 & +68 07 37.9 & 20.74 & 0.67 & 2.29 & $-0.07$ & \,$-9.52$ &  $-10.11$ & $0.31$ \\
F3.2 & 03 46 06.46 & +68 05 50.6 & 21.35 & 1.04 & 1.83 & ~~$0.45$ & \,$ -8.45$ &  $ -8.60$ & $0.11$ \\
F3.4 & 03 45 58.07 & +68 05 40.8 & 21.78 & 0.53 & 2.11 & $-0.15$ & \,$ -8.30$ &  $ -9.72$ & $0.51$ \\
F3.6 & 03 46 15.32 & +68 06 53.5 & 21.63 & 0.51 & 2.14 & $-0.18$ & \,$ -8.48$ &  $-10.27$ & $0.52$ \\
F3.7 & 03 46 12.42 & +68 04 45.8 & 21.08 & 0.96 & 2.48 & ~~$0.16$ & \,$ -9.37$ &  $ -9.48$ & $0.11$ \\
F3.8 & 03 46 02.37 & +68 05 45.6 & 21.83 & 1.07 & 2.23 & ~~$0.35$ & \,$ -8.37$ &  $ -8.49$ & $0.11$ \\
F4.1 & 03 46 33.35 & +68 01 21.4 & 20.66 & 1.12 & 2.23 & ~~$0.40$ & \,$ -9.54$ &  $ -9.67$ & $0.11$ \\
F4.2 & 03 46 34.03 & +68 00 54.9 & 20.15 & 0.92 & 2.14 & ~~$0.23$ & \,$ -9.96$ &  $-10.06$ & $0.11$ \\
F4.3 & 03 46 33.37 & +68 02 33.9 & 20.10 & 0.48 & 1.83 & $-0.11$ & \,$ -9.70$ &  $-10.65$ & $0.53$ \\
F4.5 & 03 46 34.76 & +68 03 41.8 & 20.48 & 0.74 & 1.98 & ~~$0.10$ & \,$ -9.47$ &  $ -9.60$ & $0.11$ \\
F5.1 & 03 47 01.69 & +68 01 26.5 & 20.73 & 0.56 & 1.98 & $-0.08$ & \,$ -9.22$ &  $ -9.88$ & $0.30$ \\
F5.2 & 03 47 04.85 & +68 05 17.4 & 19.83 & 0.81 & 2.14 & ~~$0.12$ & $-10.28$ &  $-10.40$ & $0.11$ \\
F5.3 & 03 46 58.05 & +68 04 12.7 & 21.34 & 0.70 & 1.98 & ~~$0.06$ & \,$ -8.61$ &  $ -8.78$ & $0.12$ \\
F5.4 & 03 46 44.43 & +68 02 56.0 & 20.81 & 0.46 & 2.14 & $-0.23$ & \,$ -9.30$ &  $-11.82$ & $0.74$ \\
F6.1 & 03 47 28.84 & +68 06 05.1 & 22.13 & 0.61 & 1.98 & $-0.03$ & \,$ -7.82$ &  $ -8.20$ & $0.20$ \\
F6.2 & 03 47 31.56 & +68 06 15.2 & 21.15 & 0.37 & 1.92 & $-0.25$ & \,$ -8.74$ &  $-11.53$ & $0.57$ \\
F6.3 & 03 47 29.79 & +68 04 34.9 & 21.82 & 0.59 & 2.23 & $-0.13$ & \,$ -8.38$ &  $ -9.55$ & $0.50$ \\
F6.4 & 03 47 15.72 & +68 03 48.9 & 20.97 & 0.99 & 2.29 & ~~$0.25$ & \,$ -9.29$ &  $ -9.39$ & $0.11$ \\
F6.5 & 03 47 19.45 & +68 04 29.2 & 22.03 & 0.68 & 2.23 & $-0.04$ & \,$ -8.17$ &  $ -8.60$ & $0.22$ \\
F7.1 & 03 47 00.08 & +68 09 53.8 & 19.48 & 0.59 & 2.05 & $-0.07$ & $-10.54$ &  $-11.15$ & $0.32$ \\
F7.3 & 03 47 14.85 & +68 08 57.3 & 19.33 & 0.67 & 2.05 & ~~$0.01$ & $-10.69$ &  $-10.96$ & $0.15$ \\
F7.4$^{a}$ & 03 47 13.62 & +68 08 54.8 & 19.20 & 0.47 & 2.05 & $-0.19$ & $-10.82$ &  $-12.75$ & $0.58$ \\
F8.2 & 03 45 50.37 & +68 03 16.7 & 21.56 & 0.74 & 2.26 & ~~$0.01$ & \,$ -8.67$ &  $ -8.93$ & $0.15$ \\
\hline
\multicolumn{9}{p{0.8\textwidth}}{$^{a}$ The source is excluded from further consideration, see text.}
\end{tabular}
\end{table*}

The Hertzsprung--Russell diagrams (CM diagrams) obtained from the performed photometry are typical diagrams for spiral galaxies, so we do not discuss their morphology or present each individual diagrams. In each of the eight fields, there are bright, relatively blue stars, among which we searched for massive stars. For the field F8, there is an HST image taken in H$\alpha$ filter, enabling photometry in three filters and construction of a colour-colour diagram \mbox{$(V-I)$\,--\,$(I-{\rm H\alpha}$)}. In the resulting diagram (Fig.~2), stars without emission in H$\alpha$ are located between the two parallel lines, and stars with emission are located above them. The dots represent stars up to $V = 24\,.\!\!^{\rm m}0$, with bright stars up to $V = 21\,.\!\!^{\rm m}5$ are marked with circles. A similar methodology was applied to identify stars with H$\alpha$ emission in the NGC\,4736 galaxy  (Tikhonov et al., 2021a).

\begin{figure*}
\includegraphics[width=11.5cm]{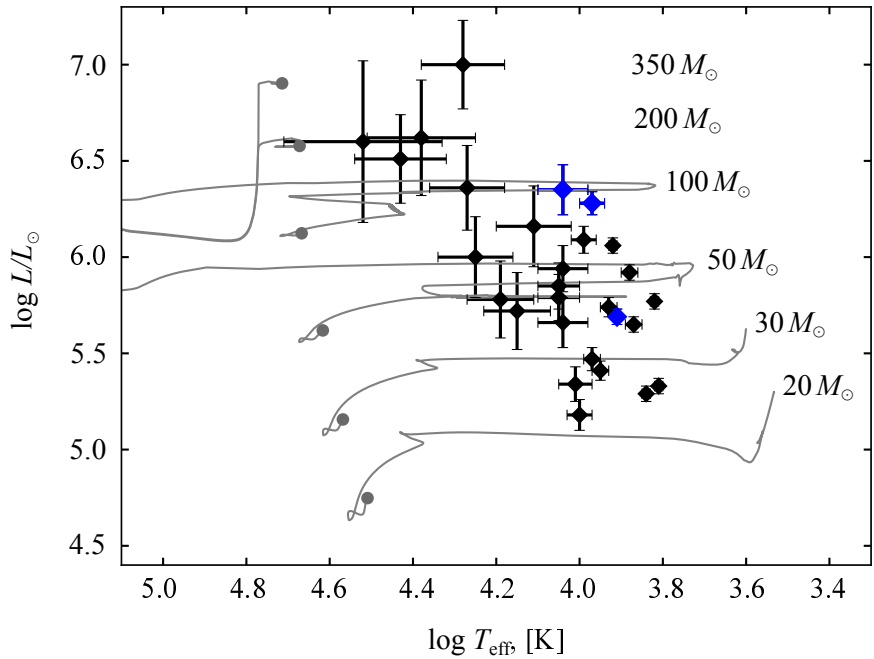}
\caption{Diagram temperature\,--\,luminosity. Blue (F3.7, F7.1 and F7.3, see text) and black (other objects) dots indicate the positions of selected sources in IC\,342 galaxy. Grey curves present the evolutionary tracks of massive non-rotating stars at $ Z=Z_{\odot}$  (Chen et al., 2015), and grey dots indicate the approximate position of the zero-age main sequence.}
\label{fig:tl}
\end{figure*}

\begin{figure*}
\includegraphics[width=450pt]{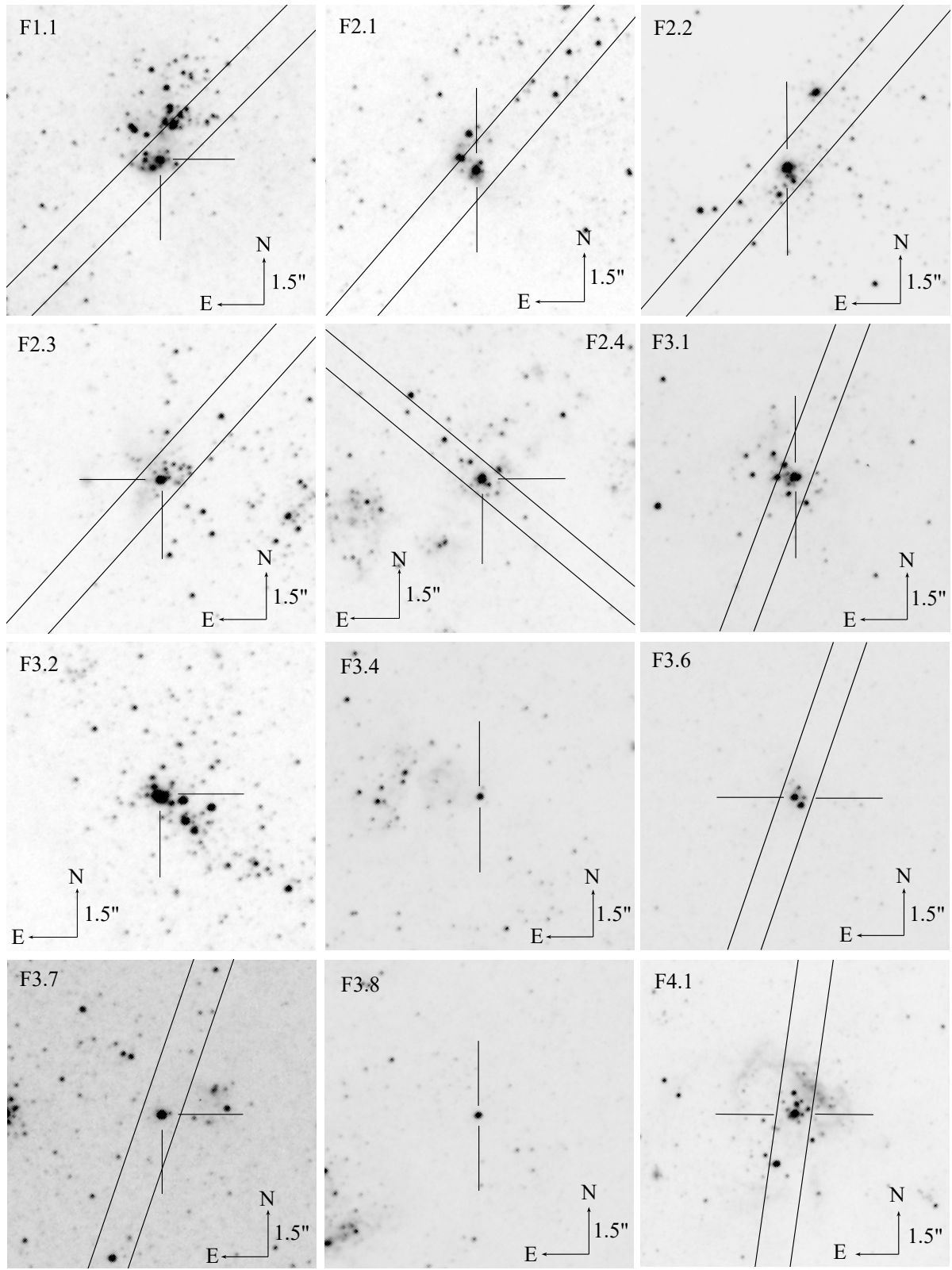}
\caption{Identification maps showing the selected candidates and the slit positions for the objects with spectroscopy.}
\label{fig:slit_1}
\end{figure*}
\setcounter{figure}{5}
\begin{figure*}
\includegraphics[width=450pt]{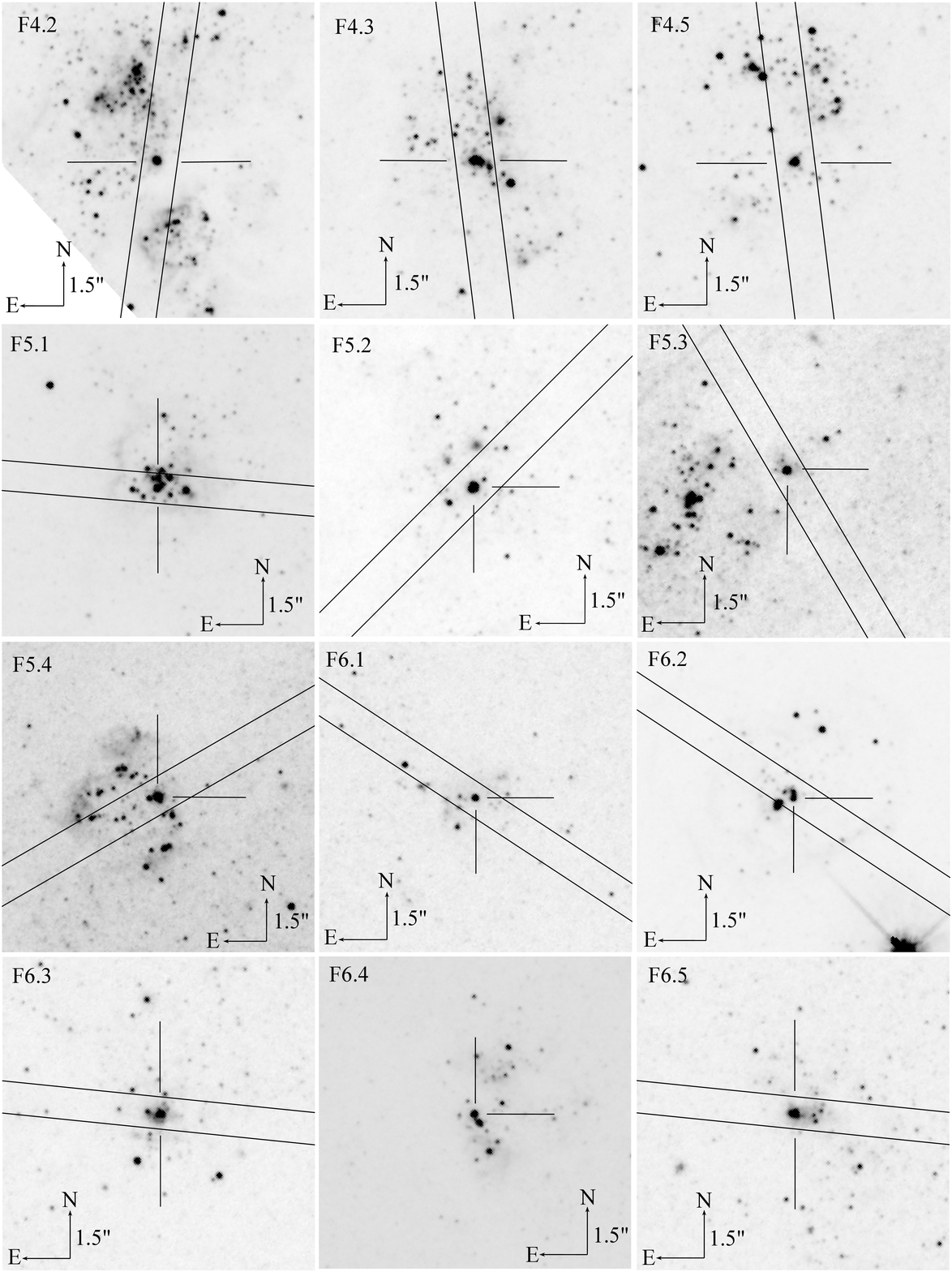}
\caption{(Contd.)}
\label{fig:slit_2}
\end{figure*}
\setcounter{figure}{5}
\begin{figure*}
\includegraphics[width=450pt]{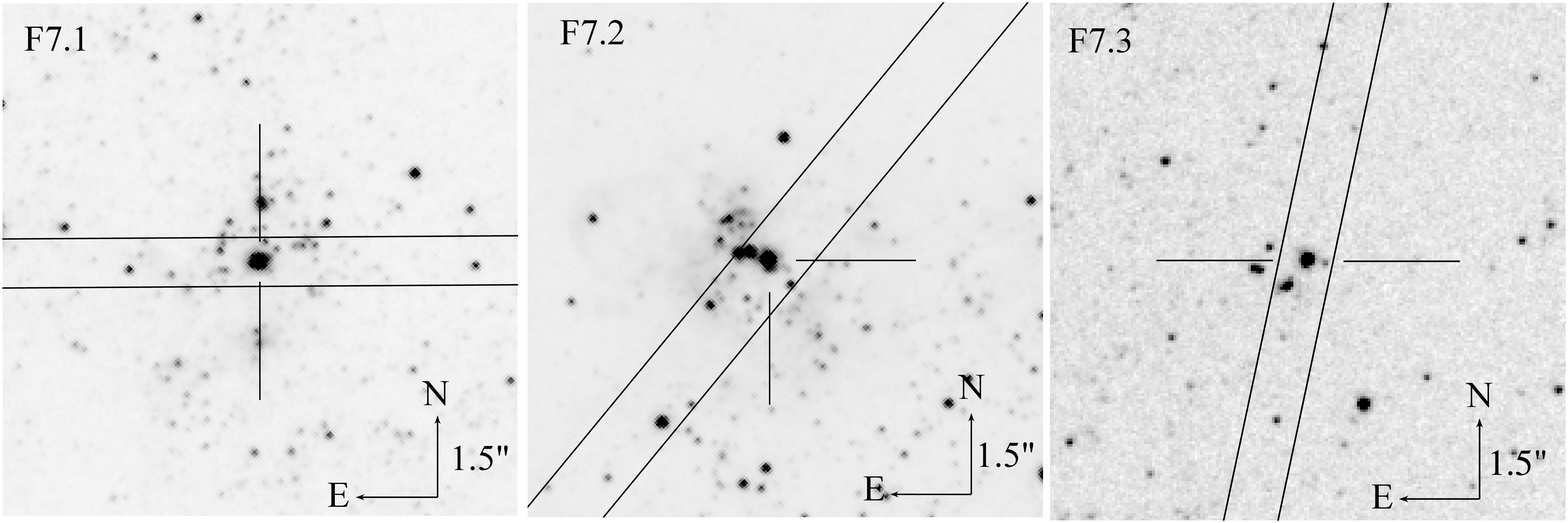}
\caption{(Contd.)}
\label{fig:slit_3}
\end{figure*}

For fields F1--F7 and F9, there are no HST images in H$\alpha$, so we changed the method of searching for massive stars. Based on $B$ and $V$ HST images, CM diagrams of the fields F1--F7 and F9 were constructed. Figure~3 shows the upper part of the general CM diagram of stars in the fields F1--F8 IC\,342 and marks the region of bright blue stars, among which massive stars should be searched. The 2~m telescope at NAO Rozhen (Bulgaria) imaged IC\,342 in $B$, $R$ and H$\alpha$ filters. Star clusters with bright blue stars found in HST images were compared with the positions of H$\alpha$ regions in images from the 2~m telescope. A bright star was considered a candidate for further investigation if its position in the HST image coincided with the position of the H$\alpha$ region in the image of the 2~m telescope. Figure~4 shows images from the 2~m telescope in $B$ and H$\alpha$ filters of F3 and F4 fields. Circles indicate clusters with bright stars and H\,{II} regions. The lifetime of the most massive stars is very short (approximately $2$--$5$\,Myr), so we excluded stellar associations with red supergiants from consideration, since they are older than 5\,Myr.

As a result, 35 sources were identified based on HST data. To exclude objects from our Galaxy from the list, we carried out cross-correlation with the Gaia\,DR3\,Part~1 catalog\footnote{\url{https://cdsarc.cds.unistra.fr/viz-bin/cat/I/355}} and found that it contained 30 objects from our sample, seven of them classified as stars with proper motion, which we excluded from further consideration. The exception was the source F5.2 in Table~1, which is also classified in the Gaia~DR3 catalog as a star with its own motion, but is identified with an X-ray source in the catalogs of Chandra and XMM observatories.

Table~1 contains selected candidates for massive stars. The numbering of the objects corresponds to the HST field numbers. The listed coordinates (RA and Dec) were determined by us on the HST ACS/WFC $F606W$ images, and the stellar magnitudes were obtained by photometry of the HST images described above. In addition, the observed colour index $(B-V)$ is given in Table~1, along with estimates of the interstellar absorption magnitude, true colour index, absolute and bolometric stellar magnitudes of the objects. When calculating the absolute stellar magnitude, the distance to the galaxy IC\,342 was assumed to be $3.93\pm0.1$~Mpc (Tikhonov and Galazutdinova,
2010, 2018)

The application of the Bertelli et al. (1994) isochrones with metallicity ${\rm Z}=0.02$ close to the metallicity of IC\,342 (Pilyugin et al., 2014) allowed us to estimate the absorptions along their lines of sight. The radius of the regions for the stars of which the measurements were made had to be chosen rather large (about 300\,pc) because of the low population of the local stellar groups closest to most of the objects in our sample. Adjusting the position of isochrones of different ages (typically from 4\,Myr to 10\,Myr), we achieved an optimal agreement of the isochrone and the blue supergiant branch. The presence on the CM-diagram of stars that had already left the blue supergiant branch allowed us to select an isochrone of the necessary age, and the shift of the isochrone along the horizontal axis indicated the amount of reddening of the colour of the blue stars. From the latter value, the extinction coefficient for a particular star cluster was calculated. The accuracy of the colour reddening is approximately $0\,.\!\!^{\rm m}02$--$0\,.\!\!^{\rm m}03$. It should be noted that not all stars from the cluster had exactly the absorption value calculated for this group of stars, because some of them are located outside the blue supergiant branch. It is impossible to draw a definite conclusion whether the deviations of these stars are consequences of their evolution stage or due to their specific extinctions.

To estimate the bolometric stellar magnitude, the effective temperatures of the objects and their corresponding bolometric corrections were determined from the synthetic photometry tables of Kurucz (Castelli, 1999)  models. Since the mean metallicity of IC\,342 is close to the solar (Pilyugin et al., 2014),  the tables corresponding to the atmospheric models with $[{\rm M}/{\rm H}]=0$ were used to determine these parameters. In this case, for simplicity, the surface gravity was set to $\log g = 4.0$, since the tabular values of synthetic colour parameters corresponding to it completely cover the range of $(B-V)_0$ values for the objects. The systematic errors caused by the uncertainty of $\log g$ do not exceed those associated with the measurement errors of the colour indexes and the interstellar absorption values. 

Based on the estimated effective temperature and bolometric luminosity, we have plotted the ``temperature\,--\,luminosity'' diagram (Fig.~5), where the selected massive star candidates are marked in black and blue (objects F3.7, F7.1, and F7.3, see below). The grey curves indicate the evolutionary tracks of massive non-rotating stars  (Chen et al., 2015),  the zero age main sequence is marked with grey dots. The diagram shows that the source F7.4 (Table~1), which has the highest luminosity estimate, does not correspond to any of the evolutionary tracks, and thus is not a single massive star. This source could be a compact young cluster that is not resolved in the HST images, so we excluded it from the list of massive star candidates in IC\,342.       

Figures~6 show the identification maps for all 27 selected candidates and the slit positions for each of the objects that were observed spectroscopically (see Sections~3--4).

\section{SPECTROSCOPY}
\label{sec:spec}

We have performed spectroscopy of 24 objects from Table~1 on the 6\,m telescope of SAO\,RAS using the SCORPIO-1 and SCORPIO-2 focal reducers in long-slit mode Afanasiev and Moiseev (2005, 2011) under the program ``Search for supermassive stars in galaxies beyond the Local Group'' led by N.~A.~Tikhonov. Some spectra were obtained under the program ``High luminosity stars in galaxies beyond the Local group'' led by Benjamin~Williams (Department of Astronomy, University of Washington) at the Apache Point Observatory (APO) 3.5~m ARC telescope using the Dual Imaging Spectrograph (DIS)\footnote{\url{https://www.apo.nmsu.edu/arc35m/Instruments/DIS/}} with a 300\,lines\,mm$^{-1}$ grating. The log of spectral observations is given in Table~2. It includes the dates of observations, instrument (or multiple instruments), grisms, exposure time $t_{\rm exp}$, slit width, resolution $R$, the spectral range $\Delta \lambda$, and image quality (seeing). All the data were processed in the {\tt Interactive Data Language} ({\tt IDL}) environment following standard techniques for processing long-slit spectra. The final extraction of spectra from the processed 2D data was performed using the program \texttt{SPEXTRA} (Sarkisyan
et al., 2017), which was specifically developed for extracting one-dimensional spectra of objects located in crowded stellar fields. 

\section{RESULTS}
\label{sec:res}

\LTcapwidth=0.95\textwidth
\setlength{\tabcolsep}{3.5pt}
\renewcommand{\baselinestretch}{0.9}
\begin{longtable*}[]{l|c|c|c|c|c|c|c}
\caption{BTA/APO Observing log} \medskip \label{tab:spec} \\
\hline
\multicolumn{1}{c|}{\multirow{2}{*}{N}} & \multicolumn{1}{c|}{\multirow{2}{*}{Date}} & \multicolumn{1}{c|}{\multirow{2}{*}{Spectrograph/Grating}} & $t_{\rm exp}$, & Slit width, & $R$,  & $\Delta \lambda$,  & \multicolumn{1}{c}{Seeing,} \\  
 &  &  & s & arcsec & \AA & \AA & arcsec \\ 
\hline 
\endfirsthead
\caption{ (Contd.) \medskip}\\
\hline
\multicolumn{1}{c|}{\multirow{2}{*}{N}} & \multicolumn{1}{c|}{\multirow{2}{*}{Date}} & \multicolumn{1}{c|}{\multirow{2}{*}{Spectrograph/Grating}} & $t_{\rm exp}$, & Slit   width,& $R$,  & $\Delta \lambda$,  & Seeing,\\  
 &  &  & s & arcsec& \AA & \AA & arcsec \\ 
\hline
\endhead
\hline
\endfoot
\hline
\endlastfoot
F1.1 & 2020/11/12--13 & SCORPIO-1/VPHG1200R & 1800 & $1.2$ & 5.0 & 5700--7400 & $1.1$ \\
F2.1 & 2022/10/25--26 & SCORPIO-1/VPHG1200R & 1800 & $1.2$ & 5.0 & 5700--7400 & $1.5$ \\ 
     & 2022/10/25--26 & SCORPIO-1/VPHG1200B & 900 & $1.2$ & 5.0 & 3600--5400 & $1.5$ \\
F2.2 & 2022/10/25--26 & SCORPIO-1/VPHG1200R & 1800 & $1.2$ & 5.0 & 5700--7400 & $1.5$ \\ 
     & 2022/10/25--26 & SCORPIO-1/VPHG1200B & 900 & $1.2$ & 5.0 & 3600--5400 & $1.5$ \\
F2.3 & 2022/12/23--24 & SCORPIO-1/VPHG1200B & 1800 & $1.2$ & 5.0 & 3600--5400 & $1.4$ \\
F2.4 & 2021/11/04--05 & SCORPIO-2/VPHG1200@540 & 1800 & $1.0$ & 5.0 & 3600--7070 & $1.3$ \\ 
    & 2022/10/25--26 & SCORPIO-1/VPHG1200R & 1800 & $1.2$ & 5.0 & 5700--7400 & $2.4$ \\ 
    & 2022/10/25--26 & SCORPIO-1/VPHG1200R & 1800 & $1.2$ & 5.0 & 5700--7400 & $2.4$ \\ 
F3.1 & 2024/01/06--07 & SCORPIO-2/VPHG1200@540 & 2700 & $1.0$ & 5.0 & 3600--7070 & $2.0$ \\
F3.6 & 2024/01/06--07 & SCORPIO-2/VPHG1200@540 & 2700 & $1.0$ & 5.0 & 3600--7070 & $2.0$ \\
F3.7 & 2020/11/13--14 & SCORPIO-1/VPHG1200R & 1800 & $1.2$ & 5.3 & 5700--7400 & $1.0$ \\ 
      & 2020/11/13--14 & SCORPIO-1/VPHG1200B & 1800 & $1.2$ & 5.5 & 3600--5400 & $1.0$ \\ 
      & 2021/10/14--15 & SCORPIO-2/VPHG1200@540 & 6300 & $1.2$ & 5.0 & 3600--7070 & $1.2$ \\ 
F4.1 & 2023/01/23--24 & SCORPIO-1/VPHG1200B & 1800 & $1.2$ & 5.0 & 3600--5400 & $1.5$ \\
F4.2 & 2023/01/23--24 & SCORPIO-1/VPHG1200B & 1800 & $1.2$ & 5.0 & 3600--5400 & $1.5$ \\
F4.3 & 2023/10/22--23 & SCORPIO-1/VPHG1200R & 1800 & $1.2$ & 5.0 & 5700--7400 & $2.0$ \\
     & 2023/10/22--23 & SCORPIO-1/VPHG1200B & 1800 & $1.2$ & 5.0 & 3600--5400 & $2.0$ \\
F4.5 & 2023/10/22--23 & SCORPIO-1/VPHG1200R & 1800 & $1.2$ & 5.0 & 5400--7400 & $2.0$ \\
     & 2023/10/22--23 & SCORPIO-1/VPHG1200B & 1800 & $1.2$ & 5.0 & 3600--5400 & $2.0$ \\
F5.1 & 2021/02/11--12 & АРО/DIS/B400 & 3600 & $0.9$ & 5.0 & 3700--5500 & $1.4$ \\     
    & 2021/02/11--12 & АРО/DIS/R300 & 3600 & $0.9$ & 5.0 & 5300--8400 &$1.4$  \\   
F5.2 & 2020/11/12--13 & SCORPIO-1/VPHG1200R & 1800 & $1.2$ & 5.0 & 5700--7400 & $1.1$ \\
      & 2022/12/01--02 & SCORPIO-2/VPHG1200@540 & 1800 & $1.0$ & 5.0 & 3600--7070 & $1.2$ \\ 
 & 2021/02/11--12 & АРО/DIS/B400 & 3600 & $0.9$ & 5.0 & 3700--5500 & $1.5$ \\     
    & 2021/02/11--12 & АРО/DIS/R300 & 3600 & $0.9$ & 5.0 & 5300--8400 &$1.5$  \\       
F5.3 & 2022/12/01--02 & SCORPIO-2/VPHG1200@540 & 1800 & $1.0$ & 5.0 & 3600--7070 & $1.2$ \\ 
F5.4 & 2023/01/26--27 & SCORPIO-1/VPHG1200R & 1800 & $1.0$ & 5.0 & 5700--7400 & $1.1$ \\
F6.1 & 2022/12/01--02 & SCORPIO-2/VPHG1200@540 & 1800 & $1.0$ & 5.0 & 3600--7070 & $1.2$ \\
F6.2 & 2022/12/01--02  & SCORPIO-2/VPHG1200@540 & 1800 & $1.0$ & 5.0 & 3600--7070 & $1.2$ \\
F6.3 & 2023/11/09--10  & SCORPIO-2/VPHG1200@540 & 1200 & $1.0$ & 5.0 & 3600--7070 & $2.0$ \\
F6.4 & 2021/02/11--12 & АРО/DIS/B400 & 3600 & $0.9$ & 5.0 & 3700--5500 & $1.6$ \\     
    & 2021/02/11--12 & АРО/DIS/R300 & 3600 & $0.9$ & 5.0 & 5300--8400 &$1.6$  \\   
F6.5 & 2023/11/09--10  & SCORPIO-2/VPHG1200@540 & 1200 & $1.0$ & 5.0 & 3600--7070 & $2.0$ \\
F7.1 & 2021/01/05--06 & APO/DIS/B400  & 1800 & $0.9$ & 5.0 & 3700--5500  & $1.4$ \\ 
     & 2021/01/05--06 & APO/DIS/R300  & 1800 & $0.9$ & 5.0 & 5300--8400  & $1.4$ \\ 
     & 2024/03/12--13 & SCORPIO-1/VPHG1200B & 2700 & $1.2$ & 5.0 & 3600--5400 & $1.5$ \\
F7.3 & 2020/11/13--14 & SCORPIO-1/VPHG1200R & 1200 & $1.2$ & 5.0 & 5700--7400 & $1.0$ \\ 
       & 2020/11/13--14 & SCORPIO-1/VPHG1200B & 1200 & $1.2$ & 5.0 & 3600--5400 & $1.0$ \\
       & 2021/11/04--05 & SCORPIO-2/VPHG1200@540 & 2700 & $1.0$ & 5.0 & 3600--7070 & $1.3$ \\
       & 2021/01/05--06 &  АРО/DIS/B400 & 3600 & $0.9$ & 5.0 & 3700--5500 & $1.4$ \\ 
       & 2021/01/05--06 & АРО/DIS/R300 & 3600 & $0.9$ & 5.0 & 5300--8400 & $1.4$ \\ 
       & 2024/03/12--13 & SCORPIO-1/VPHG1200B & 2700 & $1.2$ & 5.0 & 3600--5400 & $1.5$ \\
F8.2 & 2022/11/09--10 & SCORPIO-2/VPHG1200@540 & 3000 & $1.0$ & 5.0 & 3600--7070 & $2.1$ \\
\end{longtable*}

\begin{figure*}
\includegraphics[width=0.97\textwidth]{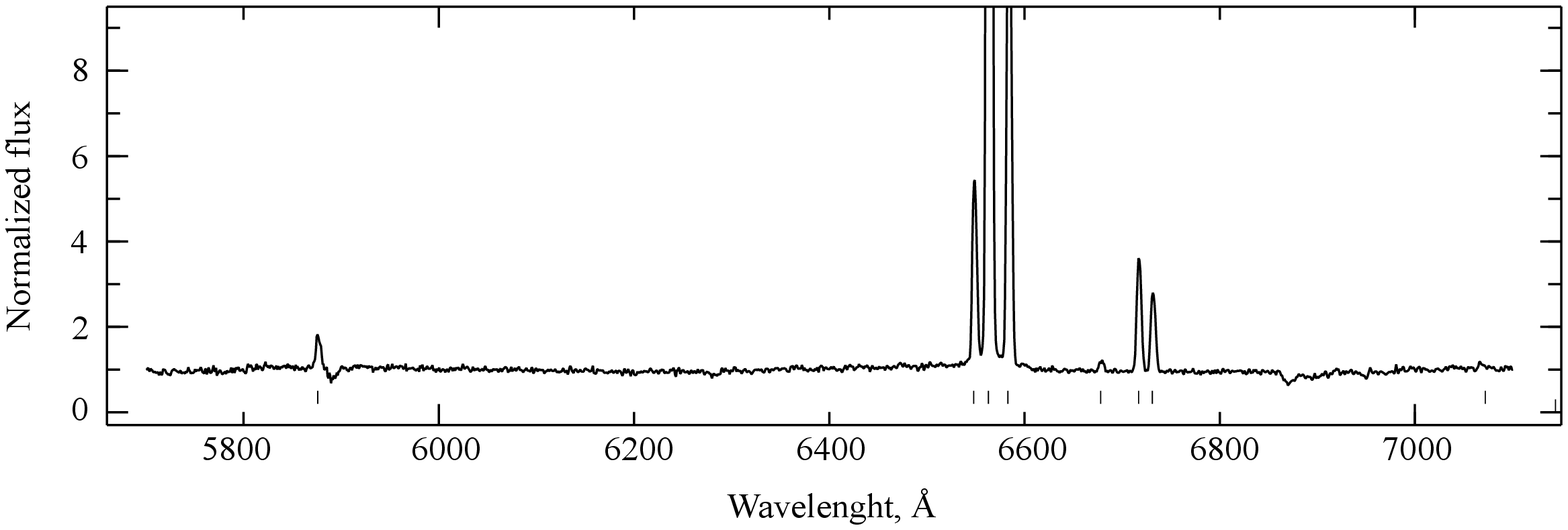}
\caption{The red region of the spectrum of object F1.1 obtained at BTA (SCORPIO-1). The nebular lines are indicated with dashes below the spectrum.}
\label{fig:sp_11R}
\end{figure*}

\begin{figure*}
\includegraphics[width=0.98\textwidth]{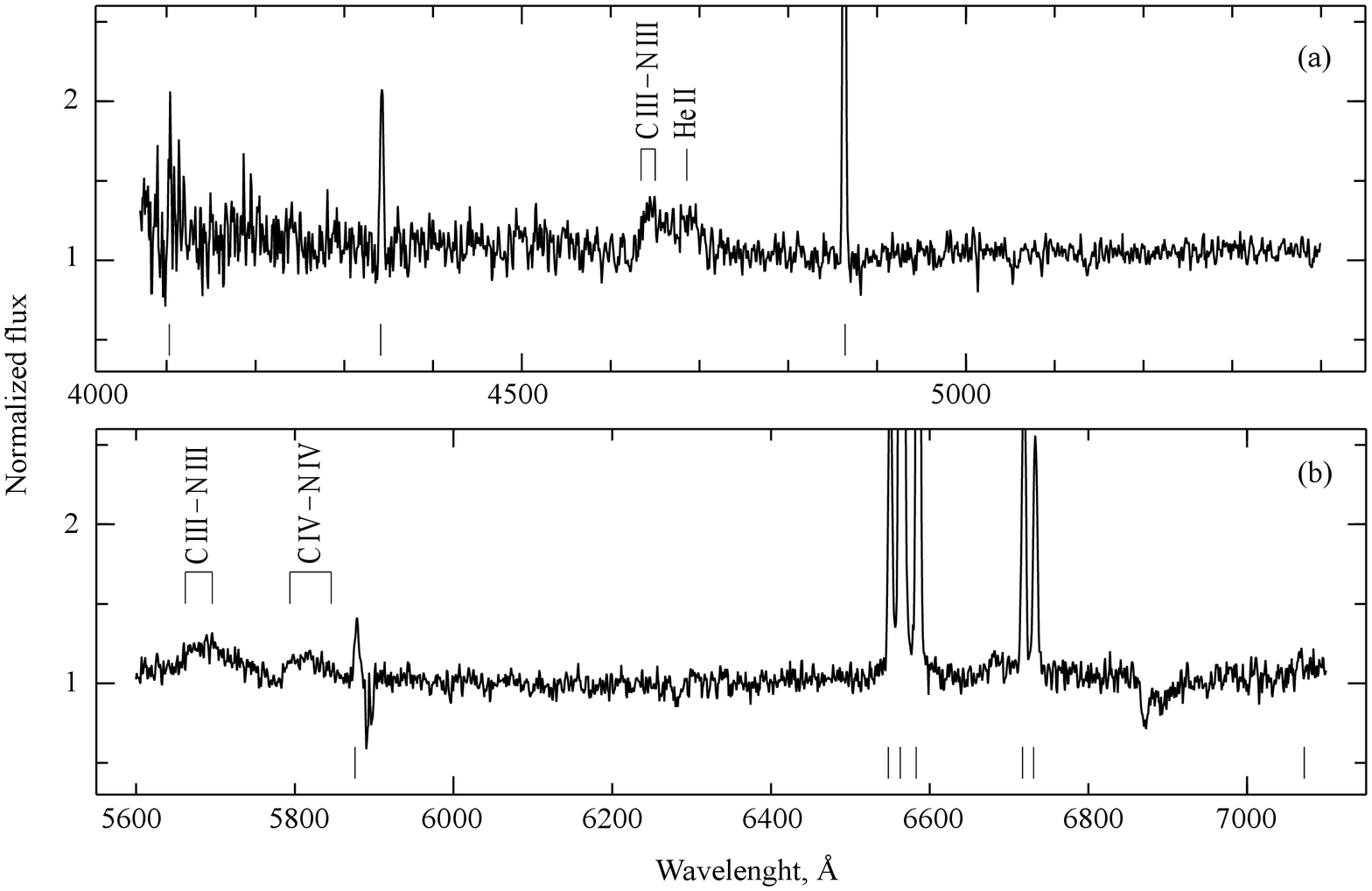}
\caption{Blue (a) and red (b) spectra of F6.1 obtained at BTA (SCORPIO-2). The nebular lines are indicated with dashes below the spectrum.}
\label{fig:sp_61R}
\end{figure*}

\begin{figure*}
\includegraphics[width=0.98\textwidth]{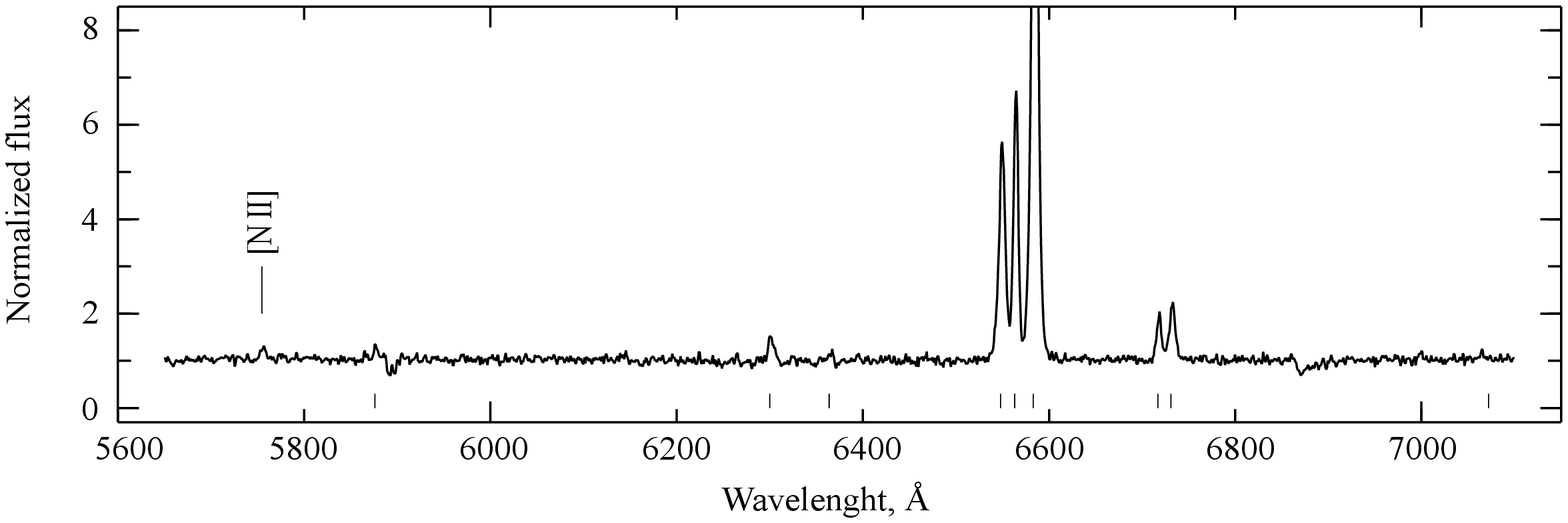}
\caption{The red region of the spectrum of object F5.2 obtained at BTA (SCORPIO-2). The nebular lines are indicated with dashes below the spectrum.}
\label{fig:sp_52R}
\end{figure*}

\begin{figure*}
\includegraphics[width=0.98\textwidth]{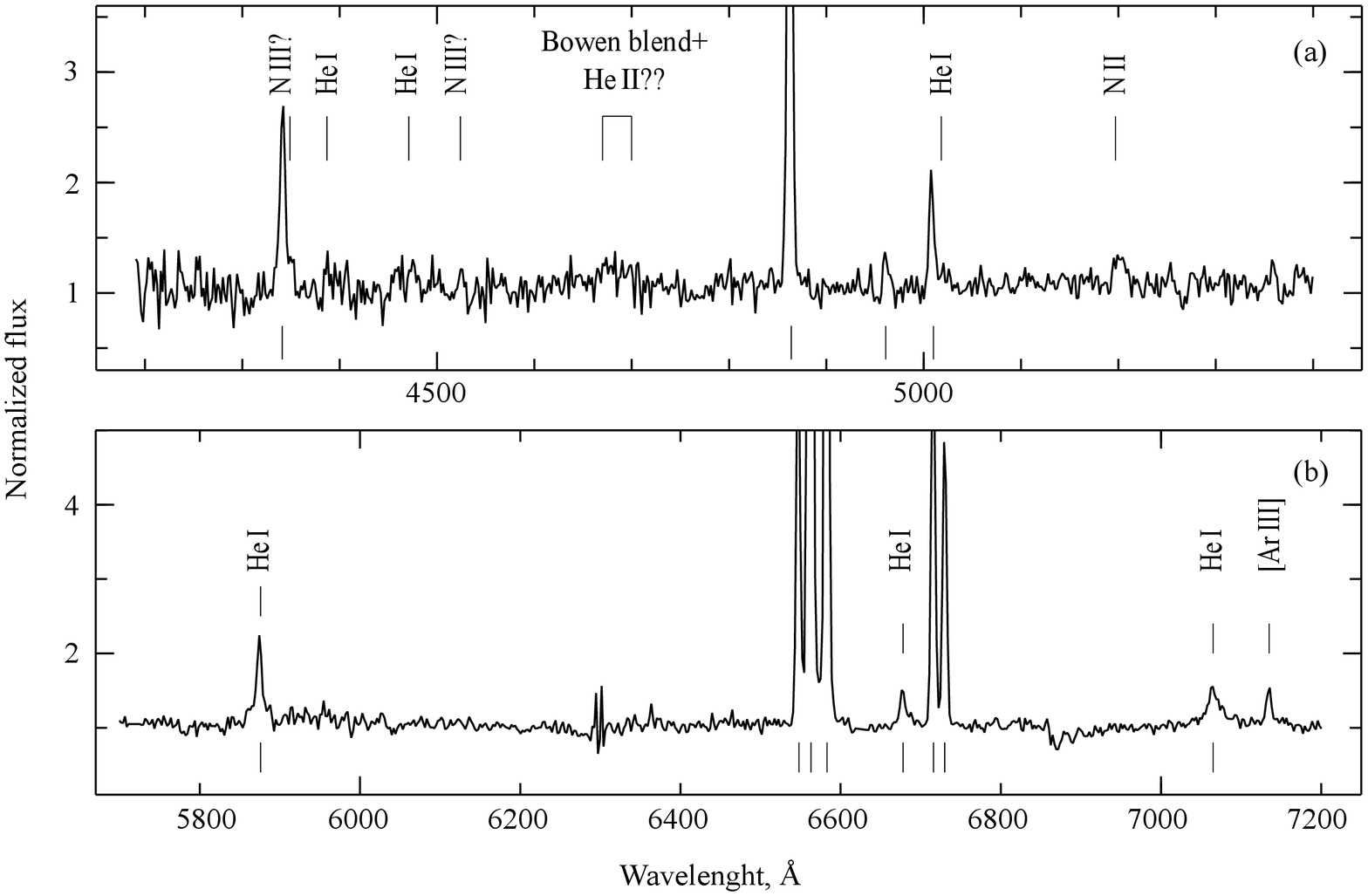}
\caption{Blue (a) and red (b) spectra of F5.1 obtained at APO (DIS). The nebular lines are indicated with dashes below the spectrum.}
\label{fig:sp_51R}
\end{figure*}

\begin{figure*}
\includegraphics[width=0.98\textwidth]{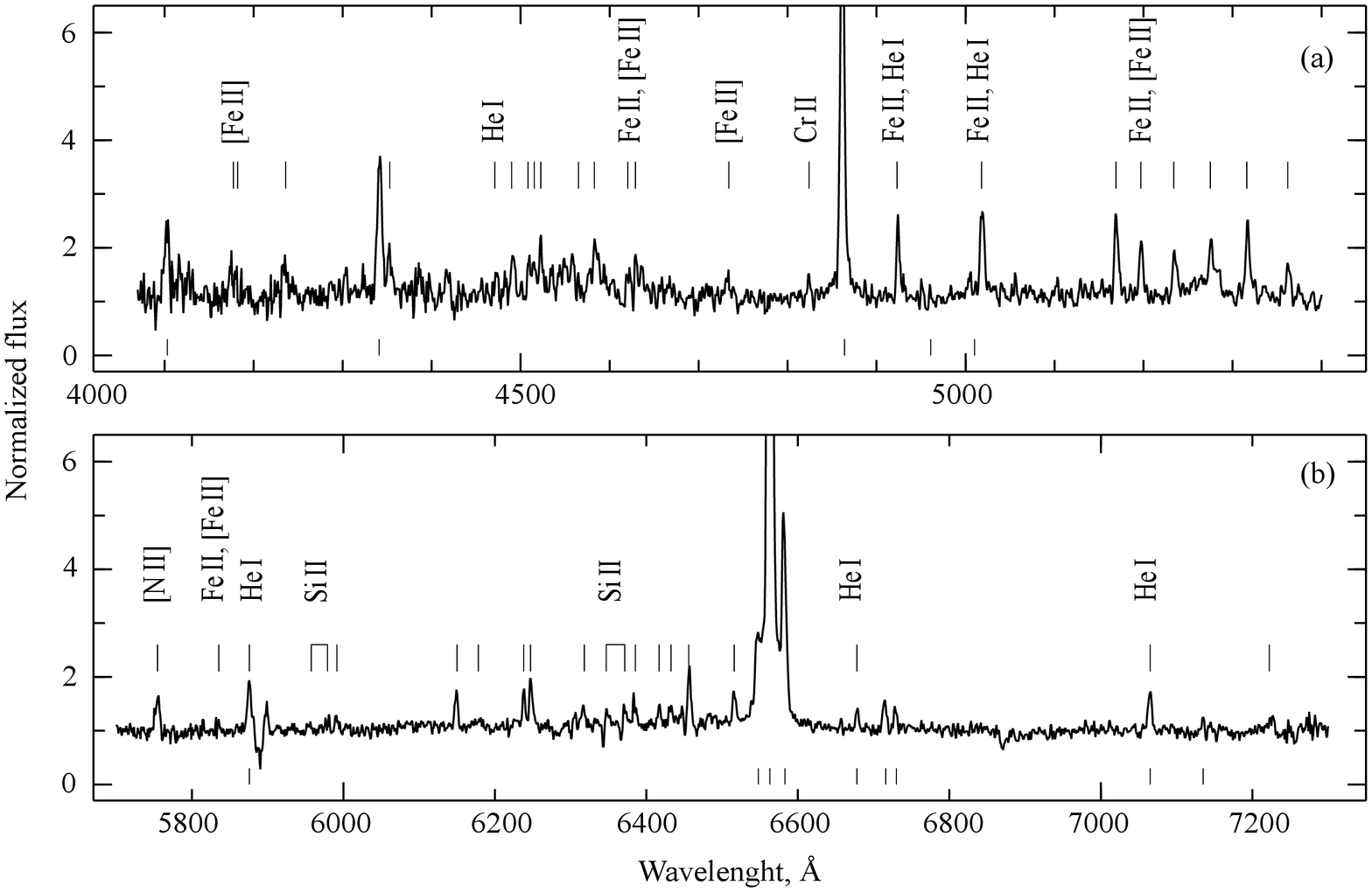}
\caption{The red region of the spectrum of object F3.7 obtained at BTA (SCORPIO-1). The nebular lines are indicated with dashes below the spectrum, Fe\,II lines---with dashes above the spectrum.}
\label{fig:sp_37B}
\end{figure*}

\begin{figure*}
\includegraphics[width=0.97\textwidth]{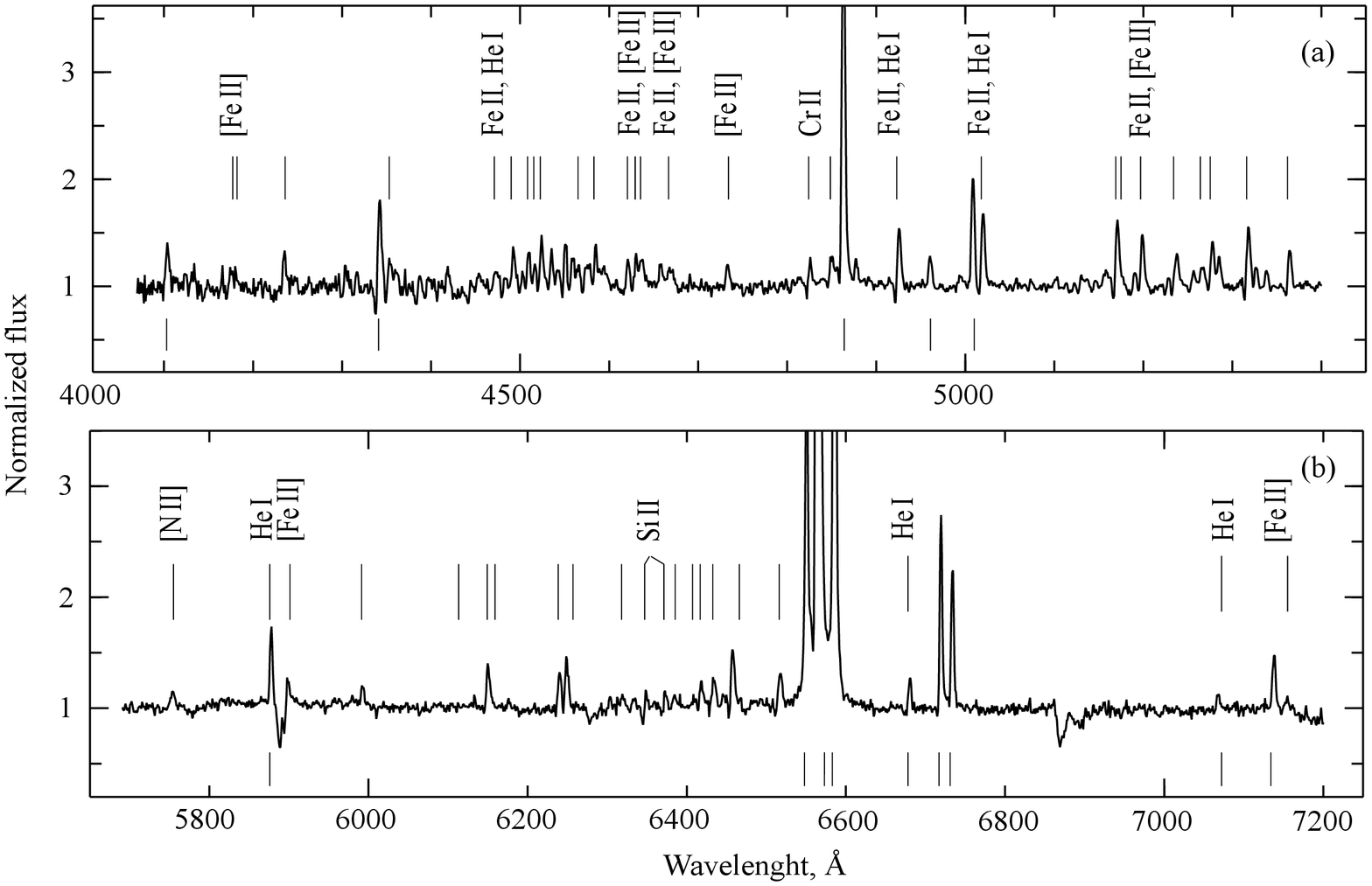}
\caption{The red region of the spectrum of object F7.3 obtained at BTA (SCORPIO-2). The nebular lines are indicated with dashes below the spectrum, Fe\,II lines---with dashes above the spectrum.}
 \label{fig:sp_73R}
\end{figure*}

\begin{figure*}
\includegraphics[width=0.98\textwidth]{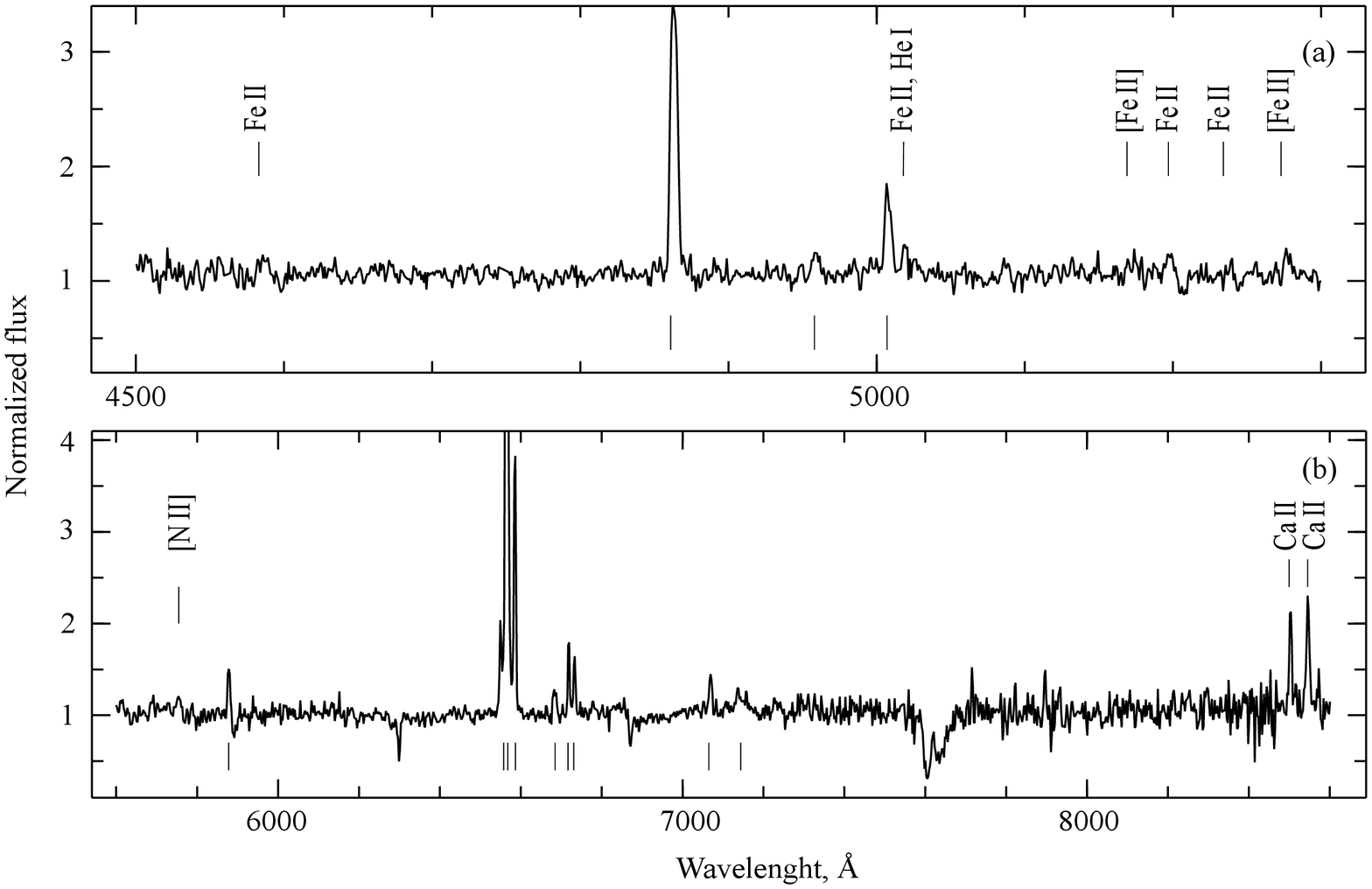}
\caption{The red region of the spectrum of object F7.1 obtained at BTA (SCORPIO-2). The nebular lines are indicated with dashes below the spectrum.}
\label{fig:sp_71}
\end{figure*}

After analyzing the spectral material for 24 objects, 12 sources (F1.1, F2.1, F3.1, F3.6, F4.5, F5.3, F5.4, F6.2, F6.4, F6.5, F8.2) were found to lack their own spectral features and contain only lines of the nebulae, which are probably not physically connected to the objects themselves, but to the whole star formation region. The brightest lines in the nebular spectra are H$\alpha$, [N\,II]\,$\lambda\,6548, \lambda\,6583$, [S\,II]\,$\lambda\,6716, \lambda\,6731$, neutral helium He\,I, oxygen [O\,III]\,$\lambda\,4959, \lambda\,5007$ and [O\,I] $\lambda\,6300, \lambda\,6330$.
The intrinsic colors $(B-V)_0$ of the listed objects correspond to spectral classes from O9 to F5 (Fitzgerald, 1970).  Figure~7 shows an example of such a spectrum of an F1.1 object in the red region.

In contrast to the previous objects, the spectrum of F2.2 is dominated by absorptions: prominent Balmer series hydrogen lines and Ca\,II, H, and K lines (the latter may be formed by the interstellar medium) are visible, as well as numerous weak, difficult to identify absorption features, some of which may be Fe\,II lines. The emission lines H$\alpha$, [N\,II]\,$\lambda\,6548, \lambda\,6583$ and [S\,II]\,$\lambda\,6716, \lambda\,6731$ belonging to the nebula are observed in the red region of the spectrum. The Gaia~DR3 catalogue originally classified the object as a galaxy with a high probability of having a reliably measured proper motion ($3.0\pm0.9$~mas\,yr$^{-1}$). At the same time, in the Gaia~DR3 catalogue of ``Quasar and galaxy classification'' (Hughes et al., 2022), this source is classified as a galaxy and a star with equal probability. Therefore, given the presence of proper motion, we do not rule out that this object is indeed a star in our Galaxy, which is projected onto the nebula-surrounded star-forming region in IC\,342.

The next group of objects (F2.3, F2.4, F4.1, F4.2, F4.3, F6.1) exhibit the following or some other spectral features inherent to \mbox{Wolf--Rayet} stars, in particular, bright blends of the C\,III\,+\,N\,III $\lambda\,4630$--$4650$ and C\,IV\,+\,N\,IV $\lambda\,5700$--$5800$, He\,II~$\lambda\,4686$ lines. As an example, the spectrum of F6.1 is shown in Fig.~8. In addition to the above lines, faint He\,I absorptions are observed in the spectra of F2.3 and F4.3. Note that the colour of these objects $(B-V)_0$ (see Table~1) is approximately \mbox{$0\,.\!\!^{\rm m}1$--$0\,.\!\!\!^{\rm m}6$} redder than the typical colours of Wolf--Rayet stars. If we assume that all objects under consideration have a non-zero colour excess due to an underestimated interstellar absorption in the region where they are located, then after an additional correction for $A_V$, the absolute stellar magnitudes of the objects will fall into the range \mbox{$M_{\rm bol}^{\rm corr} \approx -13\,.\!\!^{\rm m}0\div-14\,.\!\!^{\rm m}0$} (except for F6.1 with $M_{\rm bol}^{\rm corr} \approx -11\,.\!\! ^{\rm m}0$), that is, the luminosities of the objects would be higher than the expected values for VMS with masses 500~$M_\odot$  (e.g., Yusof
et al., 2013). Thus, it is more likely that at least the five brightest objects listed above represent young stellar associations unresolved in  HST images, which include Wolf--Rayet stars.

The observed lines in the spectrum of F5.2 (Fig.~9) are
H$\alpha$, [N\,II]\,$\lambda\,6548,6583$, [S\,II]\,$\lambda\,6716,6731$, and O\,I $\lambda\,6300,6330$, [N\,II]\,$\lambda\,5755$, and [N\,I]\,$\lambda\,5200$. Moreover, the [N\,II]\,$\lambda\,6583$ emission is brighter than the H$\alpha$ line, which is usually observed in the spectra of supernova remnants of  (e.g., Winkler et al.,
2021). It is therefore likely that F5.2 is a compact supernova remnant whose profile is not too different from the stellar PSF. Its physical  size is less than 1.4\,pc at the accepted galaxy distance of 3.9~Mpc. 

Source F5.1 is located in a rather crowded region (Fig.~6), which contributes significantly to the observed spectrum obtained when the stellar image size was $1\,.\!\!\!^{\prime\prime}6$ (Table~1). The lines of the surrounding nebula are observed in the total spectrum: the hydrogen Balmer series, [O\,III], [N\,I], [N\,II], [S\,II], [Ar\,III]), He\,I. In addition to the narrow He\,I emissions formed in the nebula, broad components are observed with $FWHM$ widths of about 1000~km\,s$^{-1}$, estimated from measurements of the He\,I\,$\lambda\,7065$ line, which appears to have the smallest contribution of the narrow component of the surrounding nebular line. At the same time, no broad components are observed in the hydrogen lines (e.g., the H$\beta$ line in Fig.~10). Also visible in the spectrum are faint lines at  $\lambda\,4348$ and $\lambda\,4524$, which may be lines of the N\,III ion, and a broad emission at $\lambda=4650$--$4700$, which is probably a blend of several lines. Overall, the F5.1 spectrum indicates that the object probably belongs to the class of late WN stars.

The most abundant emission spectra were shown by objects F3.7 and F7.3, dominated by lines of ionised iron Fe\,II and [Fe\,II] (Fig.~11 and 12). Some lines, in particular the neutral helium He\,I\,$\lambda\,6678,7065$ and Fe\,II, have P\,Cyg profiles. Both spectra also show broad components of the Balmer series hydrogen lines, few emission of Cr\,II, Si\,II, and [N\,II], with the [N\,II]\,$\lambda\,5755$ line profile having a triangular shape, which suggests line formation in wind regions with strongly differing velocities. The forbidden oxygen lines [O\,I]\,$\lambda\,6300,6363$, which, along with the forbidden [Fe\,II] lines are characteristic features of B[e]-supergiant spectra  (Humphreys et al., 2017), are not present in the F3.7 and F7.3 spectra, so both of these sources are promising LBV candidates in cold state.

\LTcapwidth=0.95\textwidth
\setlength{\tabcolsep}{5.5pt}
\begin{table}[!htp] 
\caption{Interstellar absorption values $A_V^{\rm spec}$ estimated from the hydrogen lines of the nebula around the objects. For comparison, $A_V^{\rm phot}$ derived from photometric data are given (see text) \medskip } 
\begin{tabular}{l|r@{$\,\pm\,$}l|r@{$\,\pm\,$}l} 
\hline
\multicolumn{1}{c|}{Объект} & \multicolumn{2}{c|}{$A_V^{\rm spec}$, mag} & \multicolumn{2}{c}{$A_V^{\rm phot}$, mag} \\ 
\hline
F2.3 & $2.01$ & $0.18$ & $2.36$ & $0.09$\\
F3.1 & $2.38$ & $0.13$ & $2.29$ & $0.09$\\
F3.6 & $2.87$ & $0.17$ & $2.14$ & $0.09$\\
F3.7 & \multicolumn{2}{c|}{$\approx3.6$} & $2.48$ & $0.09$\\
F4.1 & $2.19$ & $0.08$ & $2.23$ & $0.09$\\
F4.2 & $2.15$ & $0.21$ & $2.14$ & $0.09$\\
F6.2 & $2.08$ & $0.12$ & $1.92$ & $0.09$\\
F6.3 & $1.89$ & $0.28$ & $2.23$ & $0.09$\\
F6.5 & $2.87$ & $0.06$ & $2.23$ & $0.09$\\
F7.3 & $2.20$ & $0.30$ & $2.05$ & $0.09$\\
F8.2 & $2.22$ & $0.07$ & $2.26$ & $0.09$\\
\hline
\end{tabular}
\label{tab:av}
\end{table}

The spectrum of source F7.1 (Fig.~13) is similar to the spectra of F3.7 and F7.3, but has shorter total integration time. It shows broad components of hydrogen lines (H$\alpha$, H$\beta$), weak lines of ionised iron Fe\,II and possibly [Fe\,II], and bright calcium Ca\,II emission, which is part of the infrared calcium triplet at 8498~\AA, 8542~\AA, 8662~\AA{} (the latter line is outside the spectral range of our data). The presence of this triplet in stellar spectra indicates a relatively low temperature of the photosphere of the star (spectral class B–-A or later, Humphreys et al.,
2017).  Along with the forbidden calcium lines [Ca\,II]\,$\lambda\,7291,7324$, these lines are characteristic of the spectra of some B[e] supergiants and warm (yellow) hypergiants (Humphreys et al.,
2014).  In the available F7.1 spectra, only the Ca\,II lines are reliably detected, whose region is free from the contribution of atmospheric lines. Given the relatively high effective temperature of about $11\,000$~K, corresponding to a colour of $(B-V)_0 \approx -0\,.\!\!^{\rm m}07$ (see Table~1), F7.1 can be classified as a B[e] supergiant candidate, but better quality spectral data are required for its final classification. 

It was possible to estimate the magnitude of interstellar absorption from the observed ratio of the fluxes of Balmer lines of hydrogen emitted by the surrounding nebulae for 11 of the 24 objects with spectra, assuming the case of photoionisation ``B'' (Osterbrock and Ferland, 2006). The results are given in Table~3. For most sources, the Balmer decrement estimate of $A_V$ agrees within errors with the photometric-based estimate of $A_V$ (see Section~2). For three objects, the absorption values from the nebula were larger, which may indicate strong variations in absorption across the field. However, in the case of F3.7, the nebular value is clearly overestimated because the absorption-corrected colour $(B-V)_0$ is approximately $-0\,.\!\!^{\rm m}2$, which is inconsistent with the numerous Fe\,II lines and relatively weak neutral helium lines observed in the spectrum of the star. 

\section{CONCLUSION}
\label{sec:conclusions}

We have performed a selection and spectroscopic follow up study of massive star candidates in the galaxy IC\,342. We performed photometry of the stellar population of this galaxy down to $26\,.\!\!^{\rm m}5$ on archival  HST images, constructed CM-diagrams, selected bright blue stars, and compared them with the positions of H$\alpha$ emitting regions obtained from narrow-band images at the 2~m telescope of the Rozhen Observatory (Bulgaria). 

Spectra of 24 out of the 27 selected stars were obtained at the 6\,m BTA telescope of SAO\,RAN and the 3.5\,m telescope of Apache Point (USA) under a program to searching for massive stars in galaxies outside the Local Group.

Twelve objects (F1.1, F2.1, F3.1, F3.6, F4.5, F5.3, F5.4, F6.2, F6.3, F6.4, F6.5, F8.2) have no obvious features in their spectra other than bright nebular emission lines. The absence of intrinsic emission in the spectra of these stars indicates relatively weak mass loss in the form of wind. However, weak absorption lines may not be detected due to the low signal-to-noise ratio of the spectra, and even stronger hydrogen absorptions may be masked by the nebular lines mentioned above. The absolute stellar magnitudes of these sources are $M_V = -8^{\rm m}\div-10^{\rm m}$ and their true colours $(B-V)_0$ correspond to single supergiants of spectral classes O9 to F5 or to associations unresolved into individual stars. The F2.2 source, showing absorption lines along with nebular emission lines, is most likely a star in our Galaxy projected onto a star-forming region in IC\,342.

Six sources in our sample (F2.3, F2.4, F4.1, F4.2, F4.3, F6.1) have lines and blends characteristic of Wolf--Rayet stars. The five brightest sources (F2.3, F2.4, F4.1, F4.2, F4.3) appear to be unresolved young stellar associations containing Wolf--Rayet stars. Object F5.1 shows features of late Wolf--Rayet stars of the nitrogen sequence, while F5.2 is most likely a compact supernova remnant.

Two objects from our sample (F3.7 and F7.3) have spectra with features typical of cool LBV stars. Source F7.1 is probably a candidate for B[e] supergiant. For a more accurate classification of the detected objects, better quality spectra are required, as well as additional photometric data to study their luminosity. 

\section*{ACKNOWLEDGMENTS}
\label{sec:acknow}
This work is based on observations from the NASA/ESA Hubble Space Telescope obtained at the Space Telescope Science Institute, which is operated by AURA, Inc. under contract No.\,NAS5-26555. These observations are associated with applications 10768, 16002.
This work is partly based on observations obtained with the Apache Point Observatory 3.5\,m telescope, which is owned and operated by the Astrophysical Research Consortium. 
This work used NED, NED, HyperLeda and Vizier data bases.

\section*{FUNDING}
We obtained some of the observational data at the unique scientific facility Large Alt-Azimuthal Telescope of SAO\,RAS and performed the processing and analysis of the observational data under a grant from the Ministry of Science and Higher Education of the Russian Federation No.\,075-15-2022-262 (13.MNPMU.21.0003).
This study is financed by the European Union-NextGenerationEU, through the National Recovery and Resilience Plan of the Republic of Bulgaria, project No.\,BG-RRP-2.004-0008-C01 acknowledged by A.V. and P.N.

\section*{CONFLICT OF INTEREST}
The authors declare no conflict of interest.

\onecolumngrid

\begin{thebibliography}{38}
\providecommand{\natexlab}[1]{#1}

\bibitem[{Afanasiev} and {Moiseev}(2005)]{1}
V.~L.~{Afanasiev} and A.~V.~{Moiseev}, Astronomy Letters \textbf{31}, 194 (2005). DOI:10.1134/1.1883351 

\bibitem[{Afanasiev} and {Moiseev}(2011)]{2}
V.~L.~{Afanasiev} and A.~V.~{Moiseev}, Baltic Astronomy \textbf{20}, 363 (2011). DOI:10.1515/astro-2017-0305 

\bibitem[{Bertelli} et~al.(1994)]{3}
G.~{Bertelli}, A.~{Bressan}, C.~{Chiosi}, et~al., Astron. and Astrophys. Suppl.  \textbf{106}, 275 (1994). DOI:10.1051/0004-6361:20030970  

\bibitem[{Castelli}(1999)]{4}
F.~{Castelli}, Astron. and Astrophys. \textbf{346}, 564 (1999). 

\bibitem[{Chen} et~al.(2015)]{5}
Y.~{Chen}, A.~{Bressan}, L.~{Girardi}, et~al., Monthly Notices Royal Astron. Soc.  \textbf{452}~(1), 1068 (2015). DOI:10.1093/mnras/stv1281 

\bibitem[{Dolphin}(2016)]{6}
A.~{Dolphin}, {DOLPHOT: Stellar photometry}, Astrophysics Source Code Library, record ascl:1608.013 (2016). 

\bibitem[{Eikenberry} et~al.(2004)]{7}
S.~S.~{Eikenberry}, K.~{Matthews}, J.~L.~{LaVine}, et~al., Astrophys.~J. \textbf{616}~(1), 506 (2004). DOI:10.1086/422180 

\bibitem[{Figer} et~al.(2020)]{8}
D.~F.~{Figer}, F.~{Najarro}, M.~{Messineo}, et~al., Astrophys.~J. \textbf{901}~(1), id.~L15 (2020). DOI:10.3847/2041-8213/abb704  

\bibitem[{Fitzgerald}(1970)]{9}
M.~P.~{Fitzgerald}, Astron. and Astrophys. \textbf{4}, 234 (1970). 

\bibitem[{Hosokawa} and {Omukai}(2009)]{10}
T.~{Hosokawa} and K.~{Omukai}, Astrophys.~J. \textbf{703}~(2), 1810 (2009). DOI:10.1088/0004-637X/703/2/1810 

\bibitem[{Hughes} et~al.(2022)]{11}
A.~C.~N.~{Hughes}, C.~A.~L.~{Bailer-Jones}, and S.~{Jamal}, Astron. and Astrophys. \textbf{668}, id.~A99 (2022). DOI:10.1051/0004-6361/202244859 

\bibitem[{Humphreys} et~al.(2017)]{12}
R.~M.~{Humphreys}, M.~S.~{Gordon}, J.~C.~{Martin}, et~al., Astrophys.~J. \textbf{836}~(1), article id.~64 (2017). DOI:10.3847/1538-4357/aa582e 

\bibitem[{Humphreys} et~al.(2014)]{13}
R.~M.~{Humphreys}, K.~{Weis}, K.~{Davidson}, et~al., Astrophys.~J. \textbf{790}~(1), article id.~48 (2014). DOI:10.1088/0004-637X/790/1/48 

\bibitem[{Klochkova}(2019)]{14}
V.~G.~{Klochkova}, Astrophysical Bulletin \textbf{74}~(4), 475 (2019). DOI:10.1134/S1990341319040138 

\bibitem[{Kostenkov} et~al.(2017)]{15}
A.~{Kostenkov}, S.~{Fabrika}, O.~{Sholukhova}, et~al., ASP Conf. Ser., \textbf{510}, 457 (2017).

\bibitem[{Kraus}(2019)]{16}
M.~{Kraus}, Galaxies \textbf{7}~(4), id.~83 (2019). DOI:10.3390/galaxies7040083 

\bibitem[{Kurtev} et~al.(2007)]{17}
R.~{Kurtev}, J.~{Borissova}, L.~{Georgiev}, et~al., Astron. and Astrophys. \textbf{475}~(1), 209 (2007). DOI:10.1051/0004-6361:20066706 

\bibitem[{Oh} and {Kroupa}(2016)]{18}
S.~{Oh} and P.~{Kroupa}, Astron. and Astrophys. \textbf{590}, id.~A107 (2016). DOI:10.1051/0004-6361/201628233 

\bibitem[{Oskinova} et~al.(2013)]{19}
L.~M.~{Oskinova}, M.~{Steinke}, W.~R.~{Hamann}, et~al., Monthly Notices Royal Astron. Soc.  \textbf{436}~(4), 3357 (2013). DOI:10.1093/mnras/stt1817 

\bibitem[{Osterbrock} and {Ferland}(2006)]{20}
D.~E.~{Osterbrock} and G.~J.~{Ferland}, \emph{Astrophysics of gaseous nebulae and active galactic nuclei}, 2nd ed. (University Science Books, Sausalito, CA, 2006). 

\bibitem[{Pastorello} and {Fraser}(2019)]{21}
A.~{Pastorello} and M.~{Fraser}, Nature Astronomy \textbf{3}, 676 (2019). DOI:10.1038/s41550-019-0809-9  

\bibitem[{Pilyugin} et~al.(2014)]{22}
L.~S.~{Pilyugin}, E.~K.~{Grebel}, and A.~Y.~ {Kniazev}, Astron.~J.  \textbf{147}~(6), article id.~131 (2014). DOI:10.1088/0004-6256/147/6/131  

\bibitem[{Portegies Zwart} et~al.(1999)]{23}
S.~F.~{Portegies Zwart}, J.~{Makino}, S.~L.~W.~{McMillan}, and P.~{Hut}, Astron. and Astrophys. \textbf{348}, 117 (1999). DOI:10.48550/arXiv.astro-ph/9812006  

\bibitem[{Richardson} and {Mehner}(2018)]{24}
N.~D.~{Richardson} and A.~{Mehner}, Research Notes of the American Astronomical Society \textbf{2}~(3), article id.~121 (2018). DOI:10.3847/2515-5172/aad1f3  

\bibitem[{Sarkisyan} et~al.(2017)]{25}
A.~N.~{Sarkisyan}, A.~S.~{Vinokurov}, Y.~N.~{Solovieva}, et~al., Astrophysical Bulletin \textbf{72}~(4), 486 (2017).   DOI:10.1134/S1990341317040137 

\bibitem[{Sholukhova} et~al.(2018)]{26}
O.~N.~{Sholukhova}, S.~N.~{Fabrika}, A.~F.~{Valeev}, and A.~N.~{Sarkisian}, Astrophysical Bulletin \textbf{73}~(4), 413 (2018). DOI:10.1134/S199034131804003X  

\bibitem[{Solovyeva} et~al.(2019)]{27}
Y.~{Solovyeva}, A.~{Vinokurov}, S.~{Fabrika}, et~al., Monthly Notices Royal Astron. Soc.  \textbf{484}~(1), L24 (2019). DOI:10.1093/mnrasl/sly241  

\bibitem[{Solovyeva} et~al.(2021)]{28}
Y.~{Solovyeva}, A.~{Vinokurov}, A.~{Sarkisyan}, et~al., Monthly Notices Royal Astron. Soc.  \textbf{507}~(3), 4352 (2021). DOI:10.1093/mnras/stab2036  

\bibitem[{Solovyeva} et~al.(2023)]{29}
Y.~{Solovyeva}, A.~{Vinokurov}, N.~{Tikhonov}, et~al., Monthly Notices Royal Astron. Soc.  \textbf{518}~(3), 4345 (2023). DOI:10.1093/mnras/stac3408  

\bibitem[{Stetson}(1987)]{30}
P.~B.~{Stetson}, Publ. Astron. Soc. Pacific \textbf{99}, 191 (1987). DOI:10.1086/131977  

\bibitem[{Stetson}(1994)]{31}
P.~B.~{Stetson}, Publ. Astron. Soc. Pacific  \textbf{106}, 250 (1994). DOI:10.1086/133378  

\bibitem[{Tikhonov} et~al.(2021{\natexlab{a}})]{32}
N.~{Tikhonov}, O.~{Galazutdinova}, O.~{Sholukhova}, et~al., Research in Astronomy and Astrophysics \textbf{21}~(4), id.~098 (2021{\natexlab{a}}). DOI:10.1088/1674-4527/21/4/98  

\bibitem[{Tikhonov} and {Galazutdinova}(2010)]{33}
N.~A.~{Tikhonov} and O.~A.~{Galazutdinova}, Astronomy Letters \textbf{36}~(3), 167 (2010). DOI:10.1134/S1063773710030023 

\bibitem[{Tikhonov} and {Galazutdinova}(2018)]{34}
N.~A.~{Tikhonov} and O.~A.~{Galazutdinova}, Astrophysical Bulletin \textbf{73}~(3), 279 (2018). DOI:10.1134/S1990341318030021  

\bibitem[{Tikhonov} et~al.(2019)]{35}
N.~A.~{Tikhonov}, O.~A.~{Galazutdinova}, and G.~M.~{Karataeva}, Astrophysical Bulletin \textbf{74}~(3), 257 (2019). DOI:10.1134/S1990341319030027 

\bibitem[{Tikhonov} et~al.(2021{\natexlab{b}})]{36}
N.~A.~{Tikhonov}, O.~A.~{Galazutdinova}, G.~M.~{Karataeva}, et~al., Astrophysical Bulletin \textbf{76}~(4), 381 (2021{\natexlab{b}}). DOI:10.1134/S1990341321040143  

\bibitem[{Winkler} et~al.(2021)]{37}
P.~F.~{Winkler}, S.~C.~{Coffin}, W.~P.~{Blair}, et~al., Astrophys.~J. \textbf{908}~(1), id.~80 (2021). DOI:10.3847/1538-4357/abd77d  

\bibitem[{Yusof} et~al.(2013)]{38}
N.~{Yusof}, R.~{Hirschi}, G.~{Meynet}, et~al., Monthly Notices Royal Astron. Soc.  \textbf{433}~(2), 1114 (2013). DOI:10.1093/mnras/stt794 

\end{thebibliography}
\end{document}